\documentclass[aps,prb,superscriptaddress,twocolumn,longbibliography]{revtex4-2}
\usepackage{cancel}
\usepackage{physics}
\usepackage[english]{babel}
\usepackage[utf8]{inputenc}
\usepackage{amssymb}
\usepackage{amsmath}
\usepackage{multirow}
\usepackage{float}
\usepackage[pdftex]{graphicx}
\usepackage{xspace}
\usepackage{dsfont}
\usepackage{soul}
\usepackage{graphics}
\usepackage{comment}
\usepackage{svg}

\usepackage{color,xcolor}

\usepackage{hyperref}
\hypersetup{colorlinks=true,citecolor=blue,linkcolor=blue,filecolor=blue,urlcolor=blue,pdfstartview=FitH,pdfpagemode=UseNone}
\usepackage{soul} 
\usepackage[normalem]{ulem} 
\definecolor{boxcolor}{HTML}{108f64}

\begin{document}

\title{Polar unidirectional magnetotransport in $p-$type tellurene from quantum geometry}

\author{Claudio Iacovelli}
\affiliation{Independent Researcher, 08193 Barcelona, Spain}

\author{Pierpaolo Fontana}
\affiliation{Departament de Física, Universitat Autònoma de Barcelona, 08193 Bellaterra, Spain}

\author{Victor Velasco}
\affiliation{International School for Advanced Studies (SISSA), Via Bonomea 265, I-34136 Trieste, Italy}

\author{Chang Niu}
\affiliation{Elmore Family School of Electrical and Computer Engineering, Purdue University, West Lafayette, Indiana 47907, United States} 
\affiliation{Birck Nanotechnology Center, Purdue University, West Lafayette, Indiana 47907, United States}

\author{Peide~D.~Ye}
\affiliation{Elmore Family School of Electrical and Computer Engineering, Purdue University, West Lafayette, Indiana 47907, United States} 
\affiliation{Birck Nanotechnology Center, Purdue University, West Lafayette, Indiana 47907, United States}

\author{Marcus V. O. Moutinho}
\affiliation{Universidade Federal do Rio de Janeiro - Campus Duque de Caxias, Duque de Caxias, 25240-005 Rio de Janeiro, Brazil}

\author{Caio Lewenkopf}
\affiliation{Instituto de F\'isica, Universidade Federal do Rio de Janeiro, 21941-972 Rio de Janeiro, Brazil}

\author{Marcello B. Silva Neto}
\affiliation{Instituto de F\'isica, Universidade Federal do Rio de Janeiro,  21941-972 Rio de Janeiro, Brazil}

\date{\today}

\begin{abstract}
Unidirectional magnetoresistance, or electric magnetochiral anisotropy (eMChA), is a nonlinear magnetotransport phenomenon that arises in noncentrosymmetric conductors , where changes in resistance $R(B)$ are: (i) chiral, $\Delta R(B)/R(0)=2\,\chi\, {\bf I}\cdot{\bf  B}$, or (ii) polar, $\Delta R(B)/R(0)=2\,\gamma\, {\bf I}\cdot({\bf P}\times{\bf  B})$, with eMChA coefficients $\chi$ and $\gamma$. In [Phys. Rev. Lett. 135, 106602 (2025)], we showed that the eMChA in the conduction band of tellurene is polar ($\chi=0$, $\gamma\neq 0$) and emerges from the quantum metric dipole due to its Weyl node and from the lone pair polarization ${\bf P}$. Here, we extend our work to the valence band of tellurene, where the eMChA is usually said to be chiral ($\chi \neq 0, \gamma = 0$). We show that also a polar coefficient $\gamma \neq 0$ emerges naturally through a downfolding procedure, in which remote Weyl-node containing bands induce momentum-space gradients of the quantum metric in the low-energy levels, activating finite metric dipoles. Combining semiclassical Boltzmann transport with a ${\bf k}\cdot{\bf p}$ description of tellurene, our numerical calculations agree  quantitatively with doping ($\mu$) dependent second-harmonic measurements of the longitudinal voltage $V^{2\omega}_\parallel(\mu)$ in perpendicular field. The combined chiral and polar characters ($\chi\neq0, \gamma\neq 0)$ of the eMChA in tellurene also explains the shift in the angular ($\phi$) dependence of $V^{2\omega}_\parallel(\phi)$ for in plane fields. Our results demonstrate that the polar eMChA can arise in topologically trivial bands through multiband effects and establishes tellurene as a platform for quantum-geometric rectification in both electron and hole regimes.
\end{abstract}

\maketitle

\noindent

\section{\label{Intro}Introduction}
\begin{figure}[t]
    \includegraphics[trim={0pt 160pt 0pt 80pt}, clip,scale=0.29]{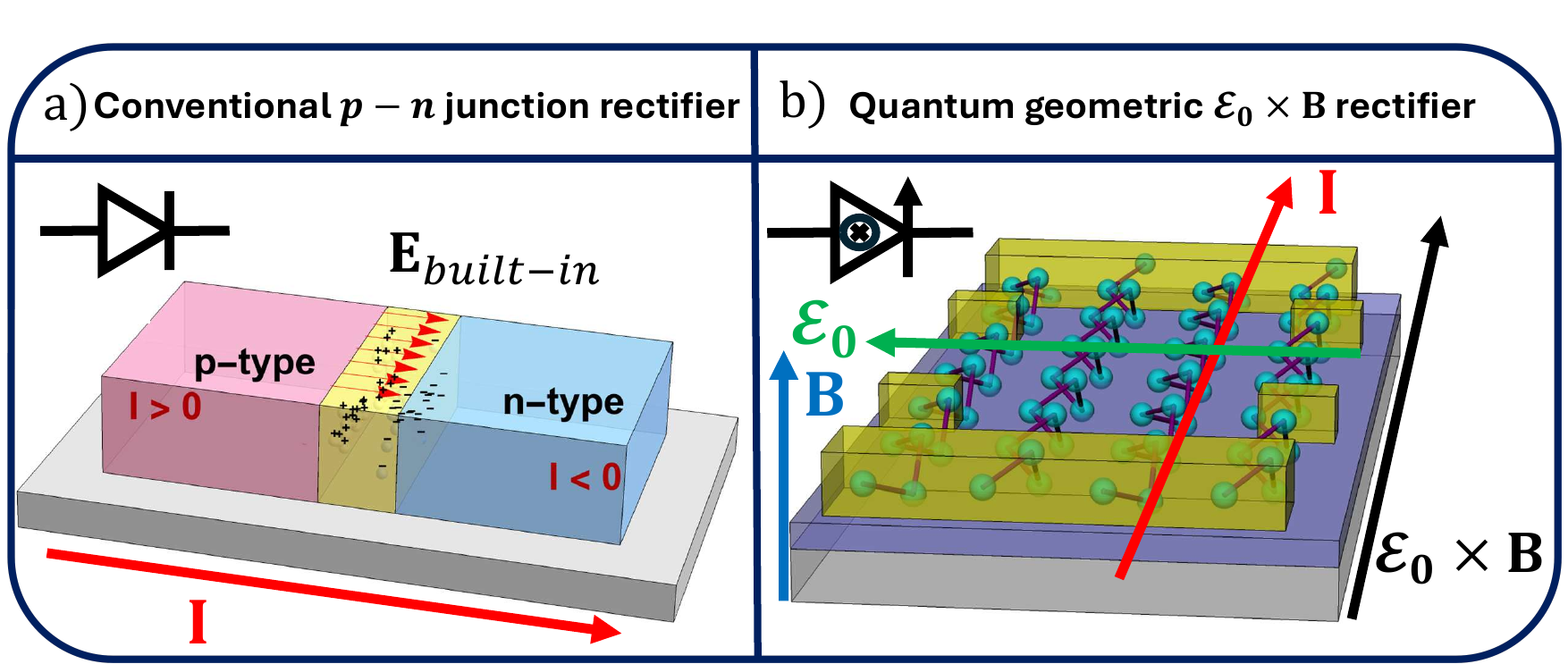}
    \caption{\textbf{Extrinsic vs intrinsic rectification.} $(\mbox{a})$ In a conventional $p$–$n$ junction diode, rectification originates from the built-in electric field ${\bf E}_{{built\text{-}in}}$ created at the interface between $p$-type and $n$-type regions. $(\mbox{b})$ In tellurene, instead, the intrinsic lone pair polarization electric field $\boldsymbol{{\cal E}}_{0}\parallel\hat{x}$ arising at the surfaces of the material plays the role of the built-in field. When an external magnetic field ${\bf B}\parallel\hat{y}$ is applied, carriers experience a nonreciprocal response along the direction $\boldsymbol{{\cal E}}_{0}\times{\bf B}\parallel\hat{z}$, contributing to the longitudinal eMChA. 
}
    \label{fig:pn_vs_me_rectification}
\end{figure}

Nonreciprocal transport phenomena occur when the flow of carriers depends on the direction of the applied current, leading to asymmetric conduction \cite{Fundamental-Electric-Circuits}. In conventional devices such as \textit{p--n} junctions, this nonreciprocity is an extrinsic effect arising from built-in electric fields at interfaces, which enables rectification and diode operation \cite{Fundamental-Power-Electronics}, see Fig.~\ref{fig:pn_vs_me_rectification}a). 
In contrast, crystalline solids lacking inversion symmetry can exhibit intrinsic nonreciprocal transport \cite{suarez2025nonlineartransport}. 
This symmetry breaking underpins a diverse range of phenomena, including natural optical activity in chiral materials \cite{NaturalOpticalActivity}, directional magnon transport in magnetic insulators \cite{Chiral-Magnons}, nonlinear phononic responses \cite{PhononicRectification}, nonlinear Hall effect \cite{GiantNLHERectification}, and electric magnetochiral anisotropy (eMChA) \cite{eMChA-Polar-Semicond}.

As a prominent manifestation of this asymmetry, eMChA is a form of unidirectional magnetoresistance (UMR) observed in noncentrosymmetric systems and has been experimentally reported in polar semiconductors \cite{PolarSemiconductor}, multiferroics \cite{MultiferroicGeMnTe}, a variety of topological materials \cite{SurfaceTINanowireHeterostructure,eChMAinTBG,eMChAinZrTe5}, and chiral tellurium \cite{Rikken_2005,Rikken_2019,suarez2025symmetryorigin}. A defining feature of eMChA is that current rectification can be tuned by external control parameters such as magnetic field, gate voltage, and polarization, offering a promising route toward electrically controllable nanoscale rectification devices \cite{Dalvin}.

In noncentrosymmetric conductors, the leading UMR generally contains two distinct symmetry-allowed invariants \cite{ReviewTokuraNagaosa}. 
In polar media, the existence of a polar axis, or a macroscopic polarization $\mathbf{P}$, allows a contribution to the resistance of the form $\Delta R_P \propto \mathbf{I}\cdot(\mathbf{P}\times\mathbf{B})$.
This term is odd under inversion, as $\mathbf{P}$ is a an inversion-odd vector, and odd under time reversal, since the magnetic field $\mathbf{B}$ is time-reversal odd. 
Here, $\mathbf{I}$ represents the carrier current. 
In chiral (enantiomorphic) media, inversion symmetry is  broken even in the absence of a polar axis. 
Chirality is instead encoded in a pseudoscalar quantity, $\chi=\pm1$, which distinguishes the two enantiomers and permits a magnetochiral contribution to the resistance of the form $\Delta R_{\chi}\propto\chi\,\mathbf{I}\cdot\mathbf{B}$, often referred to as the chiral (or magnetochiral) anisotropy. 
The two-dimensional (2D) form of tellurium (Te), known as tellurene, is of particular interest as it can realize both mechanisms: it is intrinsically chiral due to its helical crystal, and in thin films or flakes it additionally develops a sizable lone-pair polarization $\mathbf{P}$ that is approximately in-plane and perpendicular to the helix axis \cite{FontanaPRL2025}.

One of the microscopic origins of polar eMChA can be traced back to Rikken’s transport analogue of magnetoelectric optical anisotropy \cite{Rikken_2002,Rikken_2005,Rikken_2019}. Rikken initially proposed that when a current ${\bf I} \sim \langle{\bf k}\rangle$ flows perpendicularly to crossed electric $\boldsymbol{\cal E}_0$ and magnetic ${\bf B}$ fields, the magnetoresistance acquires a nonreciprocal correction of the form $\Delta R(B)/R(0)=2\,\gamma\, {\bf I}\cdot(\boldsymbol{\cal E}_0\times{\bf B})$, and originates from the additional drift velocity ${\bf v} \parallel \boldsymbol{\cal E}_0 \times {\bf B}$ acquired by the charge carriers in the laboratory frame. In the context of noncentrosymmetric nanomaterials like tellurene, the field $\boldsymbol{\mathcal{E}}_0$ is not merely an external parameter, but can represent an internal polar field arising from a net polarization ${\bf P}$, broken inversion symmetry at interfaces, or band offsets in semiconductor heterostructures \cite{Rikken_2005}. This framework highlights the potential for designing intrinsic, gate-tunable rectification at the nanoscale. More recently, a more fundamental interpretation has emerged through the lens of the quantum geometry of Bloch bands \cite{QGinCondMattReview,FontanaPRL2025}. Within this framework, the quantum metric dipole (QMD) governs the geodesic flow of carriers \cite{KaplanPRR,KaplanNature,Kaplan_2024_PRL,Gravity} and, in combination with the intrinsic lone pair polarization $\mathbf{P}$ \cite{Ferro-and-Piezo-Tellurium}, this dipole generates a nonlinear current response proportional to $\mathbf{I}\cdot(\mathbf{P}\times\mathbf{B})$ \cite{ReviewTokuraNagaosa}, see Fig.~\ref{fig:pn_vs_me_rectification}(b).
In what follows, the role of the intrinsic polarization $\mathbf{P}$ will be represented by its associated internal electric field $\boldsymbol{\mathcal{E}}_0 = \mathbf{P}/(\epsilon_0 \chi_e)$, where $\epsilon_0$ and $\chi_e$ are the vacuum permittivity and the electric susceptibility, respectively.

In a previous work \cite{FontanaPRL2025}, we demonstrated that this mechanism quantitatively accounts for the polar eMChA in the conduction bands of 2D Te, where Weyl nodes located close to the band minimum give rise to strong quantum geometric effects. The mechanism was fully characterized through magnetic-field angular rotation scans and gate-voltage dependence \cite{FontanaPRL2025}. In the present study, we extend this analysis to the valence-band sector of tellurene, which is characterized by a well-defined $z-$component of orbital moments \cite{Pancharatnam-Berry,furukawa2017NComm,FurukawaPhysRevResearch,shalygin2012current,MurakamiScientificReport,Hedgehog} that leads to a pronounced chiral-type eMChA \cite{calavalle2022chargetospinconversion,suarez2025symmetryorigin}. On the other hand, the rich $4\times 4$ structure of the valence band in tellurene, with important hybridization matrix elements between the trivial $2\times 2$ and the Weyl $2\times2$ blocks \cite{Nakao_Doi_Kamimura_I,Pancharatnam-Berry}, endow hole states with a distinct quantum geometric structure. 
This raises a fundamental question: to what extent can the polar eMChA emerge within a topologically trivial valence-band manifold, and what conditions govern its magnitude?

Although theoretical predictions of topological responses in solids are abundant, their experimental realization remains comparatively scarce \cite{BansilRMP2016}. This discrepancy often stems from the fact that the relevant topological features frequently do not coincide with the Fermi energy, leading to the common assumption that their associated responses are quenched in the metallic or doped regimes \cite{Lima2025}. In topological insulators, for example, tuning the Fermi level into the bulk gap is a prerequisite to observe protected surface states \cite{Hsieh2009,Xia2009}. Here we demonstrate that this prevailing expectation does not generally hold: even when Weyl points lie deep within the band structure, their quantum geometric signatures remain operative and can give rise to measurable nonlinear transport responses \cite{Chang_2023,Li&Niu2025}. Remarkably, we find that these nontrivial contributions persist even in regions of reciprocal space where the Berry curvature vanishes. The QMD can thus generate sizable effects in the absence of singular band crossings, revealing that the geometric origin of eMChA extends well beyond conventional topological criteria.

A very recent and thorough experimental study of eMChA in trigonal $p$-type Te 
\cite{suarez2025symmetryorigin} has highlighted, with remarkable clarity, the decisive role played by crystal symmetry, anisotropic transport, and resistivity scaling in disentangling the microscopic origin of nonlinear magnetotransport in Te. By systematically measuring both longitudinal and transverse second-harmonic responses along different crystallographic directions, and by analyzing their scaling with resistivity, that work provided compelling evidence that, in the explored regime, the dominant contribution to eMChA is extrinsic and governed by symmetry-allowed scattering processes \cite{suarez2025symmetryorigin}. Their analysis establishes an essential benchmark, that in the absence of an internal built-in electric field, the nonlinear response in Te behaves as expected for a symmetry-allowed but scattering-dominated mechanism. 

The present work is fully complementary to this picture. 
Here we show that Te flakes or films possess an additional, previously overlooked ingredient, which is a net macroscopic polarization electric field $\vec{\mathcal{E}}_{0}$ arising from lone pairs at their surfaces \cite{Ferro-and-Piezo-Tellurium}. In the presence of $\vec{\mathcal{E}}_{0}\neq 0$, the same symmetry considerations discussed in \cite{suarez2025symmetryorigin} allow for an intrinsic contribution governed not by scattering, but by the quantum geometry of the Bloch bands. In this regime, the nonlinear current is therefore controlled by the band quantum metric and its dipole moment, rather than by extrinsic resistivity scaling. Thus, our results reveal a distinct and unexpected intrinsic channel for eMChA that becomes operative precisely because of the lone-pair polarization, completing the microscopic picture of nonlinear magnetotransport in tellurene.

We present a systematic theoretical and experimental study of eMChA in the valence bands of 2D Te. We demonstrate that the quantum geometric mechanism driving this effect remains robust across carrier types and energy scales, even when the relevant singular points are energetically far from the Fermi level. Our approach employs both a full multiband description and an effective low-energy Hamiltonian for the valence sector, combined with semiclassical Boltzmann transport theory to evaluate the nonlinear magnetoconductivity \cite{Ziman,Grosso_Parravicini_2000,El-Batanouny2020}. By explicitly comparing truncated and full band models, we identify the essential roles of spin--orbit coupling, band hybridization, and macroscopic polarization in enabling a finite eMChA response. Furthermore, we analyze the scaling behavior of this nonlinear response with respect to the chemical potential, contrasting it with that obtained in the conduction-band regime \cite{FontanaPRL2025}. Altogether, our results provide a unified description of quantum geometric eMChA in 2D Te across both electron- and hole-doping regimes, establishing tellurene as a model platform for quantum geometric rectification in multiband noncentrosymmetric systems \cite{Resurrection_2022}.

This paper is organized as follows. Sections \ref{sec:model} and \ref{model_valence_bands} introduce the low-energy ${\bf k}\cdot{\bf p}$ description of the valence bands in tellurene and their relevant symmetry properties. In Sec.~\ref{subsec:polarization} we introduce the lone pair polarization and discuss its role in enabling the polar type eMChA. In Sec.~\ref{subsec:Quantum-Geometry}, we review the quantum geometry of Bloch bands, followed by the semiclassical Boltzmann framework for nonlinear magnetotransport in 
Sec.~\ref{subsec:Boltzmann}, where the QMD is formally introduced. 
Sec.~\ref{sec:Geo-Rect} details the physics of quantum geometric rectification. 
In Secs.~\ref{No-QMD-H45} and \ref{Lowdin-downfolding}, we analyze the valence-band electronic structure, demonstrating how multiband hybridization and Löwdin downfolding activate the QMD. 
Section \ref{subsec:analytic-G-tensor} provides analytic expressions for the magnetochiral conductivity and its scaling with chemical potential. 
Finally, Sections \ref{sec:numerics} and \ref{sec:Comparison-Experiments} present a comparative analysis of our numerical results and experimental second-harmonic transport measurements, followed by concluding remarks in Sec.~\ref{Conclusions}.

\section{Model and Methods}
\label{sec:model}

\subsection{\label{model_valence_bands}The valence bands of Tellurium}
\begin{figure}[t]
    \includegraphics[scale=0.27]{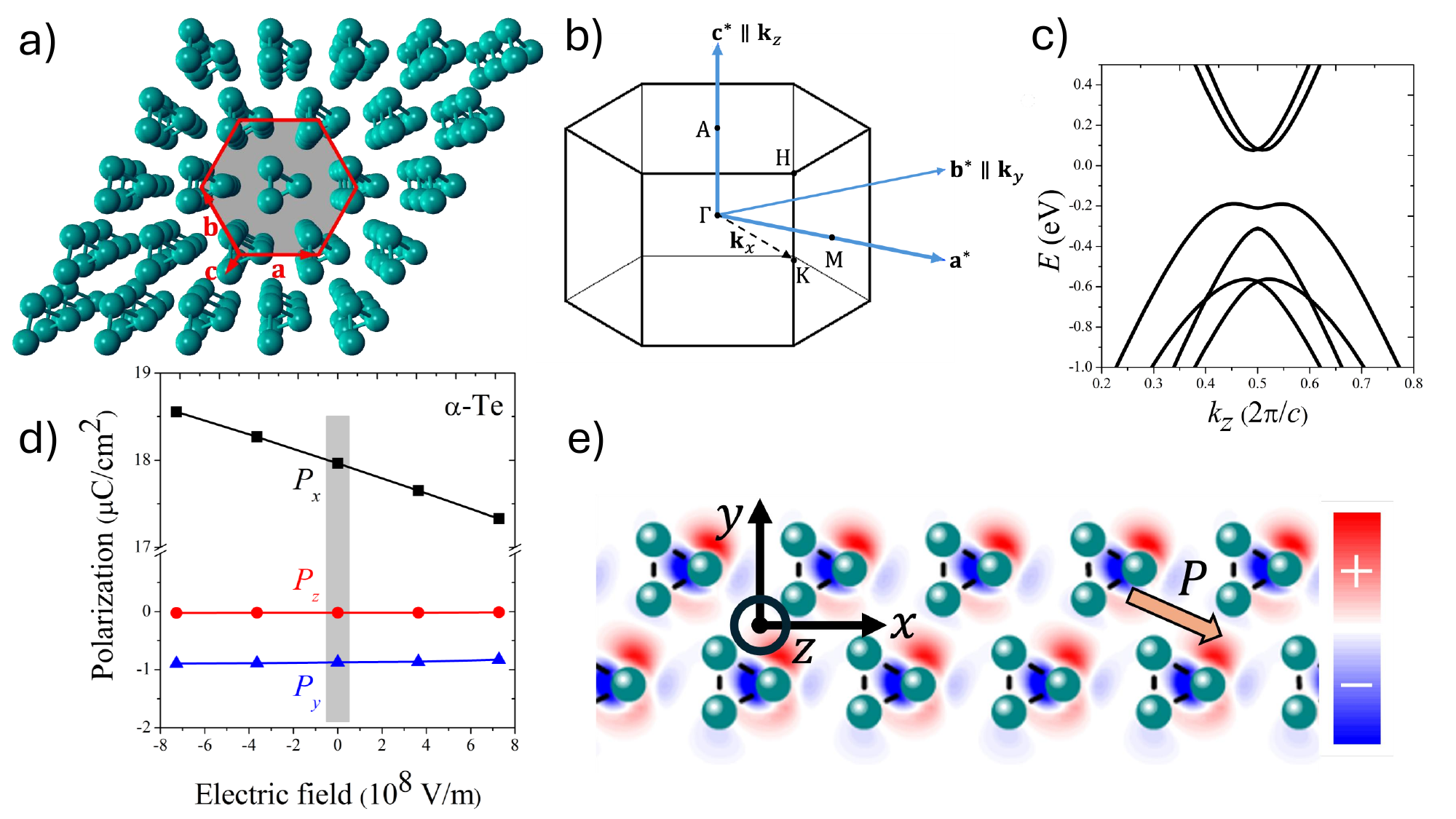}
    \caption{\textbf{Structure of tellurene.} (a) Crystal structure of $\alpha$-Te, highlighting the helical arrangement of Te atoms and the crystallographic axes. The red hexagon indicates the projection of the unit cell onto the basal plane. (b) First Brillouin zone with high-symmetry points and reciprocal lattice directions, showing the orientation of the $k_x$, $k_y$, and $k_z$ axes used throughout this work. (c) Representative band structure along the $k_z$ direction, illustrating the conduction and valence sectors relevant for nonlinear transport. (d) Calculated polarization components $(P_x,P_y,P_z)$ as a function of an applied electric field, evidencing the dominant response of the lone-pair polarization along the $x$ direction for a tellurene thin film or flake. (e) Real-space visualization of the lone-pair electronic density in a two-layer tellurene film, showing how the intrinsic dipoles align and generate a polarization $\mathbf{P}$.}
    \label{fig:crystal_bz_bands_polarization}
\end{figure}

Tellurium crystallizes in a chiral trigonal structure composed of covalently bonded helical chains that run along the crystallographic $z$ (or $c$) axis, as illustrated in Fig.~\ref{fig:crystal_bz_bands_polarization}(a). The projection of these helices onto the basal $x$-$y$ (or $a$-$b$) plane defines a natural anisotropy between in-plane and chain directions, which is directly reflected in several transport and structural anisotropies \cite{NatureOfBonding2018}. 
The valence band structure of Te can be systematically described within the framework of ${\bf k}\cdot{\bf p}$ perturbation theory \cite{Nakao_Doi_Kamimura_I,Nakao_Doi_Kamimura_II}. At the high-symmetry $H$ point of the Brillouin zone shown in Fig.~\ref{fig:crystal_bz_bands_polarization}(b), i.e., ${\bf k}_H=(4\pi/3a,0,\pi/c)$ with $a=4.44$\;\r{A} and $c=5.91$\;\r{A}, the unperturbed Hamiltonian ${\cal H}_0=\frac{{\bf p}^2}{2m}+U({\bf r})$, defines the spectrum, $E_n({\bf k}_H)$, and wave functions, $\psi_{n,{\bf k}_H}({\bf r})$, where $U({\bf r})$ is the periodic lattice potential and $n$ is the band index. To derive a description away from the $H$ point, the solutions of the full Schrödinger equation are expanded in powers of both the ${\bf k}\cdot{\bf p}$ coupling and the spin-orbit interaction (SOI) \cite{Grosso_Parravicini_2000,El-Batanouny2020}. 

Introducing Bloch states $\psi_{n,{\bf k}}(r)=e^{i{\bf k}\cdot{\bf r}}
u_{n,{\bf k}_H+ {\bf k}}
({\bf r})$, the problem is reduced to an effective eigenvalue equation for the Bloch functions $u_{n,{\bf k}_H+{\bf k}}({\bf r})$,
with energy eigenvalues $E'_n({\bf k}_H+{\bf k})=E_n({\bf k}_H+{\bf k})-\hbar^2{\bf k}^2/2m$. The corresponding Hamiltonian can be decomposed as usual $\mathcal{H}={\cal H}_0+{\cal H}_1+{\cal H}_2+{\cal H}_3$ \cite{Nakao_Doi_Kamimura_I}, with
\begin{eqnarray}
    {\cal H}_1&=&\frac{\hbar}{m}{\bf k}\cdot{\bf p},\nonumber\\
    {\cal H}_2&=&\frac{\hbar}{4m^2c^2}(\mathbf{\sigma}\times\nabla U(\mathbf{r}))\cdot\mathbf{p},\nonumber\\ 
    {\cal H}_3&=&\frac{\hbar^2}{4m^2c^2}(\mathbf{\sigma}\times\nabla U(\mathbf{r}))\cdot\mathbf{k},
\end{eqnarray}
respectively representing the ${\bf k}\cdot{\bf p}$ term (${\cal H}_1$), the ${\bf k}$-independent SOI (${\cal H}_2$) and ${\bf k}$-dependent SOI (${\cal H}_3$). All their matrix elements are evaluated with respect to the unperturbed states at the $H$ point. In the absence of SOI, the valence band state is doubly degenerate. The inclusion of SOI lifts this degeneracy, splitting the manifold into non-degenerate, irreducible valence bands, conventionally labeled as $H_4$, $H_5$, $H_6$ and $H_6'$ and shown as the four negative eigenvalues in Fig. \ref{fig:crystal_bz_bands_polarization}(c).

\begin{table}[t]
    \centering
    \begin{tabular}{l@{\hskip 1.5cm}c}
        \hline\hline
        Designation & Value \\
        \hline
        $A-r$ & $-2.67\times 10^{-15}~\text{eV cm}^{2}$ \\
        $B$ & $-3.94\times 10^{-15}~\text{eV cm}^{2}$ \\
        $S_1$ & $2.57\times 10^{-8}~\text{eV cm}$ \\
        $\Delta_1$ & $31.5\times 10^{-3}~\text{eV}$ \\
        \hline
        $A+r$ & $-6.7\times 10^{-15}~\text{eV cm}^{2}$ \\
        $S_2$ & $5.8\times 10^{-9}~\text{eV cm}$ \\
        $a$ & $\sim 10^{-8}~\text{eV cm}$ \\
        $\Delta_2$ & $-270\times 10^{-3}~\text{eV}$ \\
        \hline
        $R$ & $7\times 10^{-8}~\text{eV cm}$ \\
        $L-M$ & $3\times 10^{-14}~\text{eV cm}^{2}$ \\
        $v,G,u,b$ & $\sim 10^{-21}~\text{eV cm}^{3}$ \\
        $t,s,w_1,w_2$ & $\sim 0$ \\
        \hline\hline
    \end{tabular}
    \caption{Band parameters in the Hamiltonian \eqref{Eq: Hamiltonian-Te}, extracted from \cite{Nakao_Doi_Kamimura_II}. The separation in the three blocks is referred to the Hamiltonian blocks ${\cal H}_{45}$, ${\cal H}_{66'}$ and ${\cal T}$, from the top to the bottom of the Table.}
    \label{tab:band_params}
\end{table}

The explicit $4\times4$ block Hamiltonian in ${\bf k}$-space can be written as
\begin{equation}
{\cal H}_{4\times 4}=
\begin{pmatrix}
    {\cal H}_{45} & {\cal T} \\
    {\cal T}^\dagger & {\cal H}_{66^\prime}
\end{pmatrix}.
\label{Eq: Hamiltonian-Te}
\end{equation}
There are two $2\times2$ blocks well separated in energy. The first block, ${\cal H}_{45}$, corresponds to the bands $H_4$ and $H_5$, which are the low-energy valence bands close to the Fermi level, 
\begin{equation}
    {\cal H}_{45}=
    \begin{pmatrix}
        \varepsilon_{45}({\bf k}) - S_1 k_z - \Delta_2 & - 2\Delta_1 + u_{45}({\bf k}) \\
        - 2\Delta_1 + u^*_{45}({\bf k}) & \varepsilon_{45}({\bf k}) + S_1 k_z - \Delta_2
    \end{pmatrix},
    \label{Eq:H45}
\end{equation}
with \cite{Nakao_Doi_Kamimura_III}
\begin{eqnarray}
    \varepsilon_{45}({\bf k})&=&(A-r)k_\perp^2+B k_z^2,\nonumber\\
    u_{45}({\bf k})&=&-t k_\perp^2 + 2u k_z^2 -2iw_1 k_z,
\end{eqnarray}
where $k_\pm=k_x\pm ik_y$, $k_\perp=\sqrt{k_x^2+k_y^2}$.

The second block, ${\cal H}_{66^\prime}$, describes the bands $H_6$, $H_6'$, representing the high-energy valence bands located far from the Fermi level,
\begin{equation}
    {\cal H}_{66^\prime}=
    \begin{pmatrix}
        \varepsilon_{66^\prime}({\bf k}) + S_2 k_z + \Delta_2 & a k_+ + v_{66^\prime}({\bf k}) \\
        a k_- + v^*_{66^\prime}({\bf k}) & \varepsilon_{66^\prime}({\bf k}) - S_2 k_z + \Delta_2
    \end{pmatrix},\label{Eq: H66}
\end{equation}
with \cite{Nakao_Doi_Kamimura_III}
\begin{eqnarray}
    \varepsilon_{66^\prime}({\bf k})&=&(A+r)k_\perp^2+B k_z^2,\nonumber\\
    v_{66^\prime}({\bf k})&=&-t k_-^2 + 2i s k_+ k_z.
\end{eqnarray}
The low- and high-energy subspaces are coupled by SOI, represented by a $2\times 2$ hybridization matrix
\begin{equation}
    {\cal T}=
    \begin{pmatrix}
         (-iR+w_1)k_- + \gamma({\bf k}) & i(b^*-iw_2) k_-+\eta({\bf k})\\
         -i(b^*-iw_2) k_+ +\eta^*({\bf k}) & (iR-w_1)k_+ + \gamma^*({\bf k})
    \end{pmatrix},\label{Eq :Tau}
\end{equation}
where \cite{Nakao_Doi_Kamimura_III}
\begin{eqnarray}
    \eta({\bf k})&=&-v k_-k_z,\nonumber\\
    \gamma({\bf k})&=&\left(-i\frac{L-M}{2}-s\right)k_+^2+
    (G-iu)k_-k_z.\label{Eq :eta_gamma_k}
\end{eqnarray}
%
\begin{figure}[t]
    \includegraphics[scale=0.5]{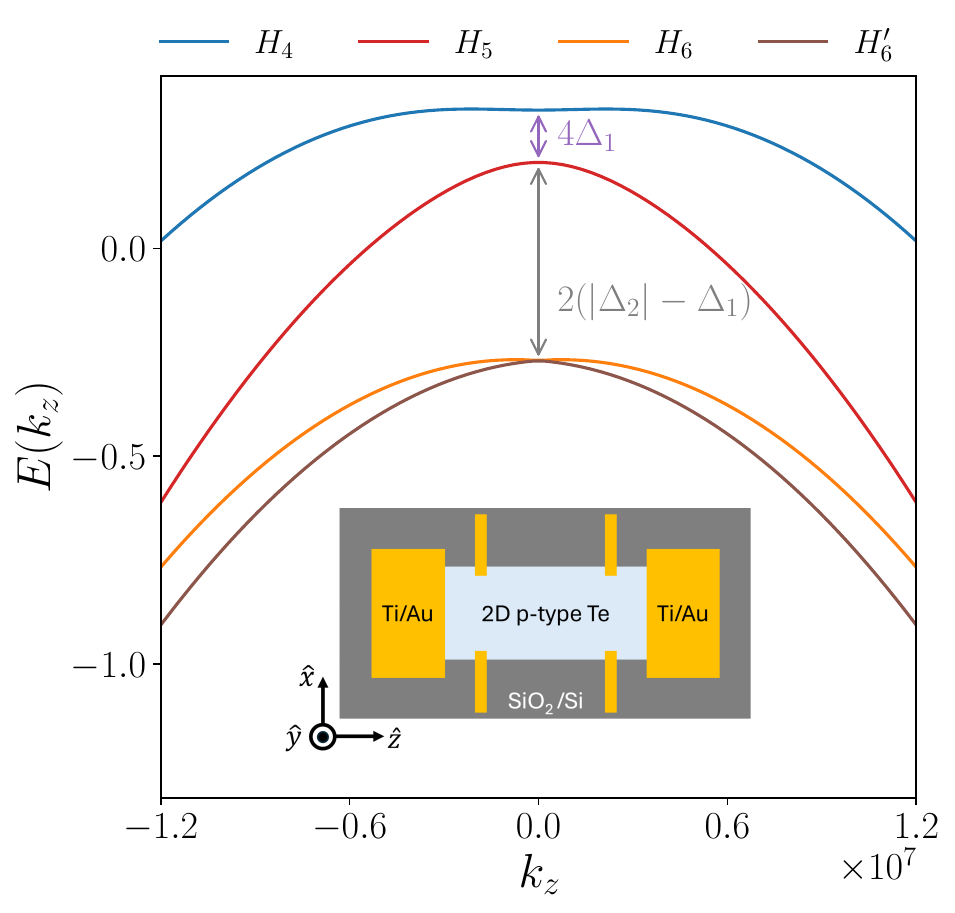}
    \caption{\textbf{Valence band structure and device geometry.}
    Valence band structure of Te along the line $k_x=k_y=0$, calculated using the Hamiltonian in Eq.~\eqref{Eq: Hamiltonian-Te} with the parameters listed in Table \ref{tab:band_params}. The energy scales $\Delta_1$ and $\Delta_2$ represent the gaps between the $H_{4,5}$ and $H_{6,6'}$ band blocks, as indicated by the purple and gray vertical arrows, respectively. Inset: Pictorial representation of the four-terminal device used for nonreciprocal transport measurements within the $x-z$ plane.}
    \label{fig:full_bands_Te}
\end{figure}
The constant parameters in all the above Hamiltonian terms come from the matrix elements of the perturbed Hamiltonian \cite{Nakao_Doi_Kamimura_I,Nakao_Doi_Kamimura_II,Nakao_Doi_Kamimura_III}. In particular, the parameters $\Delta_1$ and $\Delta_2$, arising at first order in the ${\bf k}$-independent SOI, rule the band splitting (see Fig. \ref{fig:full_bands_Te}) and are the largest scales in the system. The capital parameters, i.e. $A$, $B$, $G$, $L$, $M$, $R$, $S_1$ and $S_2$, arise at first and second order in the ${\bf k}\cdot{\bf p}$ term, while lowercase letters come either from SOI, i.e. $a$ and $b$, or from the combination between SOI with ${\bf k}\cdot{\bf p}$, i.e. $r$, $s$, $t$, $u$, $w_1$ and $w_2$ \cite{Nakao_Doi_Kamimura_I}.
Using the values extracted from Shubnikov-de Haas oscillations \cite{Resurrection_2022}, Lifshitz transitions under pressure \cite{PressureTe}, and magneto-optical transition measurements \cite{Magnetooptical}, reported for completeness in Table \ref{tab:band_params}, the resulting valence band structure of Te is shown in Fig. \ref{fig:full_bands_Te}. For the sake of simplicity we shall ignore the effects of trigonal warping that were considered in detail elsewhere \cite{FocassioPhysRevResearch}.

\subsection{Polarization electric field}\label{subsec:polarization}
The arrangement of lone-pair electronic density in trigonal Te \cite{NatureOfBonding2018} generates an intrinsic polarization in tellurene films or flakes \cite{Ferro-and-Piezo-Tellurium,GiantNLHERectification} and plays a central role in the transport phenomena discussed in this work. The important new ingredient for the eMChA introduced in \cite{FontanaPRL2025} is the polarization electric field $\boldsymbol{{\cal E}}_0$ arising exactly from the macroscopic lone-pair polarization in the 2D Te film, ${\bf P}$. The very nature of the lone pairs in Te forces this polarization to be nearly confined to the plane of the film, perpendicular to both the direction of the helices (${\bf z}$) and the growth direction (${\bf y}$) \cite{Ferro-and-Piezo-Tellurium}, so that ${\bf P}\parallel{\bf x}$. DFT calculations for a two-layer Te film, performed using the Quantum Espresso \cite{Giannozi2009} and Siesta \cite{Soler_2002} packages, verify that this requirement is indeed satisfied. The results are presented in Fig.~\ref{fig:crystal_bz_bands_polarization}(d)-(e). We find that the polarization is mostly oriented along the $x$-axis, ${\bf P}\parallel\pm\hat{\bf x}$, with a magnitude of roughly $\pm 18\, \mu C/cm^2$ at zero applied electric field, for $\alpha$-Te ($+$) and $\alpha^\prime$-Te ($-$) \cite{Ferro-and-Piezo-Tellurium}, and that it is robust under variations of an external field applied along the $x$-axis, as shown in Fig. \ref{fig:crystal_bz_bands_polarization}(d). Being a polar vector, ${\bf P}$ is odd under inversion, ${\cal I}$, so that ${\bf P}_{\alpha^\prime-Te}={\cal I}\;{\bf P}_{\alpha-Te}=-{\bf P}_{\alpha-Te}$, fully consistent with our DFT calculations. 

In Ref. \cite{FontanaPRL2025} we showed that, in the conduction band of tellurene, the large lone-pair polarization ${\bf P}\parallel\hat{x}$ governs the angular dependence of the unidirectional magnetoresistance, $\Delta R_{zz}({\bf B})/R_{\\}(0)=2\,\gamma^\pm\, {\bf I}\cdot({\bf P}\times{\bf B})$, when {\bf B} is confined to the $y$-$z$ and $y$-$x$ planes. Here $\gamma^\pm$ is a chirality-dependent eMChA coefficient that encodes the topological charge ($\pm$) of the Weyl node in the conduction band. We also observed a small lone-pair polarization component perpendicular to the film, ${\bf P}\parallel\mp\hat{\bf y}$, consistent with the angular variation measured for {\bf B} confined to the $x$-$z$ plane \cite{FontanaPRL2025}, while the component along the coil direction is always zero, as shown in Fig. \ref{fig:crystal_bz_bands_polarization}(e). 

\subsection{Quantum geometry}
\label{subsec:Quantum-Geometry}
Within quantum mechanics, the geometry of Bloch states is encoded in the \textit{quantum geometric tensor} (QGT) \cite{Provost1980, TopandGeoAspectsBandTheory, Kang2024QGT}. The QGT quantifies the distance in reciprocal space between neighboring Bloch states $\ket{n_{\bf k}}$ and $\ket{n_{{\bf k}+d{\bf k}}}$ of the $n$th band and is defined as
\begin{align*}
    {\cal \bf Q}^n({\bf k})
    &=
    \langle \nabla_{{\bf k}} n_{\bf k} |
    \big(1 -  |n_{\bf k}\rangle\langle n_{\bf k}|\big)
    | \nabla_{{\bf k}} n_{\bf k} \rangle ,
\end{align*}
where $\nabla_{\bf k}$ denotes the gradient in momentum space. The QGT naturally decomposes into a real, symmetric part ${\bf g}^n({\bf k})$, known as the \textit{quantum metric}, and an imaginary, antisymmetric part given by the \textit{Berry curvature} ${\bf \Omega}^n({\bf k})$, namely
\begin{align}
    \label{eq:QGT}
    {\cal \bf Q}^n({\bf k}) &=
    {\bf g}^n({\bf k})
    -\frac{i}{2}\,{\bf \Omega}^n({\bf k}).
\end{align}

While the Berry curvature ${\bf \Omega}^n({\bf k})$ acts as an effective momentum-space magnetic field, governing linear-response phenomena such as the anomalous Hall effect \cite{XiaoRMP} and topological magnetoelectric responses \cite{Hedgehog}, the quantum metric ${\bf g}^n({\bf k})$ characterizes the ``stiffness'' of the eigenstates under momentum variations. Although the quantum metric does contribute to linear transport at leading order and has recently been demonstrated to produce a giant negative magnetoresistance in tellurene \cite{GiantNMRQuantumGeometry}, it has emerged as the fundamental quantity controlling nonlinear responses \cite{QGinCondMattReview}. These include the nonlinear Hall effect, nonreciprocal magnetoresistance, and second-harmonic generation \cite{KaplanNature, Kaplan_2024_PRL}. More broadly, the quantum metric imposes geometric constraints on Bloch electrons, forcing semiclassical wave packets to follow momentum-space geodesics \cite{Gravity}, thereby modifying carrier dynamics beyond the usual Berry-curvature corrections \cite{BinghaiYan2025}.

For an effective two-band Hamiltonian of the form
\begin{equation}
    H({\bf k}) = {\bf d}({\bf k})\cdot{\boldsymbol \sigma},
    \label{eq:two_level}
\end{equation}
with energies $\varepsilon_{\pm}=\pm|{\bf d}|$, quantum geometric quantities acquire particularly transparent closed-form expressions \cite{QGinCondMattReview}. Defining $\partial_i\equiv\partial_{k_i}$, the Berry curvature of the lower band reads
\begin{equation}
    \Omega_{ij}^n
    =
    \frac{(-1)^n}{2|{\bf d}|^{3}}
    \,{\bf d}\cdot
    \left(
    \partial_i{\bf d}\times\partial_j{\bf d}
    \right),
    \label{eq:Berry_two_level}
\end{equation}
which highlights its interpretation as the flux of a momentum-space monopole associated with band degeneracies \cite{XiaoRMP}. In contrast, the quantum metric is given by
\begin{equation}
    g_{ij}^n
    =
    \frac{1}{4|{\bf d}|^{2}}
    \left[
    \partial_i{\bf d}\cdot\partial_j{\bf d}
    -
    \frac{
    (\partial_i{\bf d}\cdot{\bf d})
    (\partial_j{\bf d}\cdot{\bf d})
    }{|{\bf d}|^{2}}
    \right].
    \label{eq:metric_two_level}
\end{equation}
Unlike the Berry curvature, which requires non-coplanar variations of the $\mathbf{d}$ vector to remain nonzero, the quantum metric depends only on the magnitude and local gradients of $\mathbf{d}$. Consequently, the metric can remain finite even in regions of the Brillouin zone where the Berry curvature vanishes. This distinction is central to the present work, as it enables nontrivial geometric responses in bands that are otherwise topologically trivial.

To describe nonlinear transport phenomena sensitive to interband virtual processes, it is convenient to introduce an energy-weighted version of the quantum geometric tensor \cite{Kaplan_2024_PRL}. This is obtained by scaling each virtual transition in Eq.~\eqref{eq:QGT} by the corresponding interband energy denominator $\varepsilon^n_{\bf k}-\varepsilon^m_{\bf k}$ \cite{Kaplan_2024_PRL}. We denote the real part of the rescaled tensor as $\tilde{\bf g}^n({\bf k})$, which, for two-level systems, reduces to $\tilde g_{ij}=g_{ij}/(2|{\bf d}|)$.

The momentum-space curvature of this rescaled metric is captured by the Christoffel symbols \cite{Gravity}
\begin{equation}
    \Gamma^{n}_{ij\ell}
    =
    \frac{1}{2}
    \left(
    \partial_i \tilde g^{n}_{j\ell}
    +
    \partial_j \tilde g^{n}_{i\ell}
    -
    \partial_\ell \tilde g^{n}_{ij}
    \right),
    \label{eq:Christoffel}
\end{equation}
where the explicit $\mathbf{k}$-dependence is omitted for brevity. These symbols encode the {\it dipolar structure} of the quantum metric across the Brillouin zone. Just as the Berry curvature gradient (the Berry curvature dipole) drives nonlinear Hall currents, these Christoffel symbols govern the nonlinear corrections to semiclassical dynamics that give rise to unidirectional transport \cite{FontanaPRL2025}. In the following, we demonstrate that the emergence of finite Christoffel symbols serves as a sharp criterion for the emergence of eMChA in the valence-band sector of tellurene.

\subsection{{Semiclassical transport and quantum metric}}
\label{subsec:Boltzmann}
To evaluate the nonreciprocal response of hole carriers in 2D Te, we employ the semiclassical Boltzmann transport theory \cite{Ziman,Grosso_Parravicini_2000}, augmented by quantum geometric effects. This approach establishes a direct connection between the microscopic band topology and the macroscopic eMChA. In this framework, the current density is given by ($V$ is the volume of the unit cell)
\begin{equation}
    {\bf j}=-\frac{e}{V}\sum_{n,{\bf k}}\dot{\bf r}_n\,f_n({\bf k}),
    \label{eq.current}
\end{equation}
where $\dot{\mathbf{r}}_n$ represents the semiclassical velocity including geometric terms and $f_n({\bf k})$ is the non-equilibrium distribution function of the $n$-band. For a stationary and spatially homogeneous system ($\partial f_n/\partial{\bf r}=\partial f_n/\partial t=0$) and within the relaxation time approximation, the distribution function satisfies the Boltzmann equation \cite{Ziman}
\begin{equation}
    \dot{\bf k}_n\cdot\nabla_{\bf k} f_n
    =-\frac{f_n-f_n^0}{\tau}.
\end{equation}
Here, $f_n^0$ denotes the equilibrium Fermi-Dirac distribution, and $\tau$ is the relaxation time, assumed to be momentum independent (see discussion in Ref. \cite{FontanaPRL2025}).

A wave packet constructed from Bloch states of the band $n$ evolves according to the semiclassical equations of motion \cite{Chuu2010}
\begin{align}
    \dot{\mathbf{r}}_n &=\mathbf{v}_{n\mathbf{k}} + \dot{\mathbf{k}}_n\times\boldsymbol{\Omega}_{n}(\mathbf{k}) + \dot{\mathbf{k}}_n\cdot\hbar\boldsymbol{\Gamma}^{n}(\partial {\bf g})\cdot\dot{\mathbf{k}}_n,
    \label{eq:r_dot}\\[3pt]
    \hbar\dot{\mathbf{k}}_n &= -e\bigl(\mathbf{E}+\boldsymbol{\cal E}_{0}\bigr)
    - e\,\dot{\mathbf{r}}_n\times\mathbf{B},\label{Eq: Boltzmann}
\end{align}
where $\mathbf{v}_{n\mathbf{k}}=(1/\hbar)\nabla_{\mathbf{k}}\varepsilon_{n\mathbf{k}}$ is the group velocity. The second term in Eq.~\eqref{eq:r_dot} recovers the standard anomalous velocity \cite{XiaoRMP}, while the third term embodies the geodesic contribution generated by the QMD \cite{Gravity}, which is essential for nonreciprocal transport. Finally, in Eq.~\eqref{Eq: Boltzmann} $\boldsymbol{\cal E}_{0}$ represents the lone-pair polarization electric field.

In the DC limit $\omega\tau\ll 1$ and under a uniform static magnetic field ${\bf B}$, the nonequilibrium distribution function $f_n$ acquires a correction due to the Lorentz force, which is given by 
\begin{equation}
    \delta f_n
    =e\tau\!\left[(\dot{\bf r}_n\!\cdot\!{\bf E})
    +e\tau\!\left(\frac{\dot{\bf r}_n\times{\bf B}}{\hbar}\right)\!\cdot\!
    \nabla_{\bf k}(\dot{\bf r}_n\!\cdot\!{\bf E})\right]
    \left(\frac{\partial f^0_n}{\partial\varepsilon_n}\right).
\end{equation}
By substituting the modified velocity $\dot{\mathbf{r}}_n$ from Eq.~\eqref{eq:r_dot} into the expression for the current density and isolating the contribution arising from the QMD, which depend on ${\bf \Gamma}^n(\partial{\bf g})$, we find a nonreciprocal current at order $E^2B$ \cite{Spivak_2023}
\begin{equation}
    \delta{\bf j}=\frac{2e^5\tau^2}{\hbar^2 V}\sum_{n,{\bf k}}{\bf v}^n_{\bf k}
    [({\bf E}\cdot{\bf \Gamma}^n(\partial{\bf g})\cdot\boldsymbol{\cal E}_0)\times{\bf B}]\cdot\nabla_{\bf k}({\bf v}^n_{\bf k}\cdot{\bf E})
    \left(-\frac{\partial f^0_n}{\partial \varepsilon_n}\right).
\end{equation}
This expression has the structure $\langle{\bf v}_{\bf k}\rangle_{\rm QMD}\!\cdot\!(\boldsymbol{\cal E}_0\times{\bf B})$, consistent with symmetry-based expectations for magnetochiral transport \cite{Rikken_2005}.

By expressing the nonreciprocal current in the tensorial form $\delta j_{i}=G_{ijk\ell}\,E_{j}E_{k}B_{\ell}$, we identify the eMChA conductivity $G_{ijk\ell}$ tensor as \cite{FontanaPRL2025}
\begin{equation}
    G_{ijk\ell}=\frac{2e^5\tau^2}{\hbar^2 V}
    \sum_{n,{\bf k}}{v}^n_{{\bf k},i}
    \left\{\epsilon_{ab\ell}{\Gamma}^n_{bkm}(\partial{\bf g}){\cal E}_{0,m}\right\}\,\partial_a {v}^n_{{\bf k},j}
    \left(-\frac{\partial f^0_n}{\partial \varepsilon_n}\right).
    \label{Eq:G-tensor}
\end{equation}
The magnitude of this response is determined entirely by the interplay between the band velocity and the Christoffel symbols of the energy-weighted quantum metric. Crucially, the integrand's symmetry dictates the vanishing or emergence of the effect: since $v_i$ and $\Gamma_{bkm}$ are both odd under inversion ($\mathbf{k} \to -\mathbf{k}$), their product is even, allowing for a non-vanishing integral. Consequently, $G_{ijk\ell}$ is non-zero only in noncentrosymmetric systems where a macroscopic polarization $\boldsymbol{\mathcal{E}}_0$ or a specific chiral arrangement is present, such as in 2D tellurene \cite{FontanaPRL2025}. 

The tensor $G_{ijk\ell}$ governs the chirality–dependent contribution to the nonreciprocal resistance. In the presence of a macroscopic polarization $\mathbf{P}$, the longitudinal resistance acquires the form (see Appendix \ref{Ap: RandGamma})
\begin{equation}
    R(\mathbf{I},\mathbf{P},\mathbf{B})
    =R_{0}\Bigl[1+\beta B^{2}+\gamma_{\rm eMChA}\,\mathbf{I}\!\cdot\!(\mathbf{P}\times\mathbf{B})\Bigr],
    \label{Eq:Rgamma}
\end{equation}
where $R_0$ is the zero-field resistance and $\beta$ is the conventional quadratic magnetoresistance coefficient \cite{GiantNMRQuantumGeometry}.
The term $\gamma_{\rm eMChA}$ quantifies the magnetochiral anisotropy, essentially measuring the strength of the nonreciprocity driven by the QMD. As detailed in Appendix~\ref{Ap: RandGamma}, this coefficient is directly accessible via second harmonic voltage measurements, where the ratio of the second to the first harmonic signals follows $V^{2\omega}/V^\omega\propto \gamma_{\rm eMChA}$.

Within the semiclassical framework and for the experimental geometry considered here (see inset of Fig. \ref{fig:full_bands_Te}), the magnetochiral coefficient $\gamma_{\rm eMChA}$ is related to the tensor $G_{ijk\ell}$ as (see Appendix~\ref{Ap: GtoGamma})
\begin{align}
    \gamma_{\rm eMChA} \propto \frac{G_{zzzy}}{\sigma_{zz}}
    \label{Eq:Gamma},
\end{align}
with the linear conductivity
\begin{eqnarray}
    \sigma_{zz}=\frac{e^2\tau}{4\pi\hbar^3 V}\sum_{n,\mathbf{k}}\left(\frac{\partial\varepsilon_n}{\partial k_z}\right)^2\frac{\partial 
    f^0_n}
{\partial\varepsilon_n}.\label{Eq:sigma_zz}
\end{eqnarray}
Thus, Eq.~(\ref{Eq:Rgamma}) provides a direct experimental manifestation of the quantum geometric tensor.
By measuring the nonreciprocal resistance, one can probe the underlying momentum-space geometry of the Bloch bands as encapsulated in the tensor response of Eq.~\eqref{Eq:G-tensor}.

\section{Geometric Rectification: $k$-dependent Stark effect.}
\label{sec:Geo-Rect}
The physical origin of the nonreciprocal current encoded in the tensor $G_{ijk\ell}$ admits a transparent interpretation as a momentum-dependent Stark renormalization of the band dispersion. In the presence of an electric field $\mathbf E$, the electric-dipole coupling can be eliminated by a similarity transformation \cite{Kaplan_2024_PRL}. This yields a quadratic correction to the band energy 
\begin{equation}
    \tilde\varepsilon_n(\mathbf k)=\varepsilon_n(\mathbf k)
    +e^2\,E_{i}E_{j}\,\tilde g^{n}_{ij}(\mathbf k),
    \label{eq:Stark_metric}
\end{equation}
where $\tilde g^{n}_{ij}(\mathbf k)$ is the energy-weighted quantum metric introduced in Sec.~\ref{subsec:Boltzmann}. Eq.~\eqref{eq:Stark_metric} reveals that the Stark effect in a crystalline solid is not merely a feature of the band dispersion, but also governed by the quantum geometry of Bloch states \cite{Kaplan_2024_PRL}.

Since the quantum metric is an even function of momentum, $\tilde g^{n}_{ij}(\mathbf k)=\tilde g^{n}_{ij}(-\mathbf k)$, the corresponding quadratic energy shift preserves inversion symmetry in momentum space. 
This property is illustrated in Fig.~\ref{fig:QM_and_Christoffel_VB}(a), where the two Fermi surfaces of the valence bands are overlaid on a heatmap of the quantum metric in the $(k_x,k_z)$ plane. The metric texture is symmetric under $k_z\rightarrow -k_z$, confirming that the Stark-induced renormalization, while shifting the local energy levels, does not by itself break the reciprocity of carrier flow.

The situation changes qualitatively when considering the semiclassical velocity. Since the band velocity is defined as the momentum derivative of the dressed dispersion, $\tilde{\mathbf v}_n=(1/\hbar)\nabla_{\mathbf k}\tilde\varepsilon_n$, the quadratic Stark correction in Eq.~\eqref{eq:Stark_metric} produces an additional velocity component
\begin{equation}
    \delta v_{n,a}(\mathbf k)\propto
    e^2\,E_{i}E_{j}\,\partial_{k_a}\tilde g^{n}_{ij}(\mathbf k),
    \label{eq:velocity_metric}
\end{equation}
which is governed by the momentum gradients of the quantum metric \cite{Gravity,Kaplan_2024_PRL}.
These gradients are naturally encoded in the Christoffel symbols $\Gamma^{n}_{ij\ell}(\partial \tilde{\bf g})$ defined in Eq.~\eqref{eq:Christoffel}, which enter the semiclassical equations of motion through the geodesic term $\dot{\mathbf k}\cdot\hbar{\bf \Gamma}^{n}\cdot\dot{\mathbf k}$ in Eq.~\eqref{eq:r_dot}.

Importantly, while the quantum metric itself is even in $\mathbf k$, its derivatives are {\it odd}, such that the Christoffel symbols satisfy
$\Gamma^{n}_{ij\ell}(\mathbf k)=-\Gamma^{n}_{ij\ell}(-\mathbf k)$.
This antisymmetry is explicitly shown in Fig.~\ref{fig:QM_and_Christoffel_VB}(b-c), for a representative Christoffel component. As a result, states at opposite regions of the Fermi surface, $\mathbf{k}$ and $-\mathbf{k}$, acquire velocity corrections of equal magnitude but opposite sign.

In the presence of a magnetic field, the Lorentz force induces a circulating motion in momentum space, causing wave packets to sample different regions of the asymmetric Christoffel texture depending on the direction of the current. This mechanism converts the even-in-$\mathbf k$ Stark energy shift into an odd-in-current velocity response, ultimately producing the rectified contribution to the current captured by the eMChA tensor $G_{ijk\ell}$ in Eq.~\eqref{Eq:G-tensor}. The nonreciprocal transport thus emerges from the interplay of the $k$-dependent Stark effect, the quantum geometry of Bloch states, and the Lorentz-induced mixing of momentum-space trajectories.

\subsection{Absence of metric dipoles in the truncated $H_4-H_5$ valence-band block}
\label{No-QMD-H45}
It is instructive to contrast the conduction-band mechanism discussed in Ref.~\cite{FontanaPRL2025} with that of the uppermost valence-band sector. To this end, we first examine the truncated $2\times2$ block spanned by the $H_{4}$ and $H_{5}$ states, as defined in Sec.~\ref{model_valence_bands}. Although this restricted subspace is endowed with a finite quantum metric, we show below that the corresponding QMDs--encoded in the Christoffel symbols relevant for eMChA--vanish identically. This demonstrates that the nonreciprocal response cannot arise within this reduced model and instead requires the full multiband structure of the valence band, including the Weyl sector formed by the $H_{6}$ and $H_{6'}$ bands.

The effective Hamiltonian for the $H_{4}$--$H_{5}$ subspace, Eq.~\eqref{Eq:H45}, can be written in terms of Pauli matrices as
\begin{equation}
    \mathcal{H}_{45}(\mathbf k)=d_0(\mathbf k)\,\mathbb{I}
    +\mathbf d_{45}(\mathbf k)\cdot\boldsymbol\sigma,
\end{equation}
where 
\begin{align}
    d_0({\bf k}) = &\, (A-r) k_\perp^2 + Bk_z^2 - \Delta_2, \nonumber\\
    \mathbf d_{45}(\mathbf k)= &\,
    \bigl(
    -2\Delta_1-tk_\perp^2+2u k_z^2,\;
    2w_1 k_z,\;
    - S_1 k_z
    \bigr).
\end{align}
For a two-level Hamiltonian of this form, the quantum metric of either band is determined by Eq.~\eqref{eq:metric_two_level}. Evaluating this expression for the $\mathcal{H}_{45}$ block, we find a finite longitudinal component
\begin{equation}
    g_{zz}^{(45)}(\mathbf k)=
    \frac{S_1^2\,
    4\Delta_1^2}
    {4\left[
    4\Delta_1^2
    +
    S_1^2k_z^2
    \right]^2},
    \label{eq:gzz_45}
\end{equation}
where we have set $t=u=0=w_1$ (see Table \ref{tab:band_params}) to focus on the leading-order geometric contributions. This quantity is manifestly even in $k_z$, as required by time-reversal symmetry, and confirms that the truncated valence-band block possesses a non-vanishing quantum metric.

In the parameter regime relevant for the $H_{4}$--$H_{5}$ sector, the pseudospin vector $\mathbf d_{45}(\mathbf k)$ depends solely on $k_z$. This occurs because the coefficients governing the $k_\perp$ dependence of the off-diagonal terms are negligible (see Table \ref{tab:band_params}). As a consequence, the quantum metric $\tilde g_{ij}$ has no dependence on $k_x$, and we find
\begin{equation}
    \partial_{k_x}\tilde g_{zz}=0,
    \qquad
    \partial_{k_x}\tilde g_{xz}=0.
\end{equation}
It follows immediately that the corresponding Christoffel symbols vanish,
\begin{equation}
    \Gamma_{xzx}^{(45)}(\mathbf k)=0,
    \qquad
    \Gamma_{zzx}^{(45)}(\mathbf k)=0.
    \label{eq:Gamma_vanish_45}
\end{equation}
This demonstrates that the uppermost valence-band sector, when treated as an isolated two-band system, is insufficient to generate any magnetochiral anisotropy.

\subsection{L\"owdin downfolding of the $H_6,\;H_{6'}$ sector and emergence of finite Christoffel symbols}
\label{Lowdin-downfolding}
To elucidate the geometric origin of the nonreciprocal response, we examine the full $4\times4$ valence-band Hamiltonian near the $H$ point, which is written according to Sec.~\ref{model_valence_bands} through Eqs.~\eqref{Eq: Hamiltonian-Te}, \eqref{Eq:H45}, \eqref{Eq: H66}, and \eqref{Eq :Tau}. Next, we employ a Löwdin downfolding procedure to derive an effective two-band model for the low-energy subspace \cite{Lowdin}. This analytical limit provides a transparent view of how the perturbative coupling to the remote $H_6$ and $H_{6'}$ sectors ``dresses" the low-energy states, thereby generating the nontrivial quantum metric and the associated Christoffel symbols required for rectification. The detailed derivation of the effective Hamiltonian is provided in Appendix~\ref{Ap: Lowdin_Appendix}. 

Given that $|\Delta_2|$ constitutes the dominant energy scale in the system (see Table~\ref{tab:band_params}), and noting that the $\mathcal{H}_{66'}$ block resides far from the Fermi level, we may treat this remote sector as dispersionless,
\begin{equation}
    \mathcal{H}_{66'}(\mathbf k)\simeq \Delta_2\,\mathbb{I}_{2\times2}.
    \label{eq:H66_dispersionless}
\end{equation}
Focusing at energies where $|E|\ll |\Delta_2|$, the L\"owdin procedure gives
\begin{eqnarray}
    \mathcal{H}_{45}^{\mathrm{eff}}(\mathbf k)
    &\simeq&
    \mathcal{H}_{45}(\mathbf k)+\mathcal{T}(\mathbf k)\,\bigl(E-\mathcal{H}_{66'}\bigr)^{-1}\mathcal{T}^\dagger(\mathbf k)
    \nonumber\\
    &\simeq& 
    \mathcal{H}_{45}(\mathbf k)-\frac{1}{\Delta_2}\,\mathcal{T}(\mathbf k)\mathcal{T}^\dagger(\mathbf k),
    \label{eq:Heff_general}
\end{eqnarray}
where to leading-order $(E-\Delta_2)^{-1}\simeq -\Delta_2^{-1}$. To highlight the novel momentum dependence introduced by the remote bands, we simplify $\mathcal{T}(\mathbf k)$ by retaining only the dominant terms listed in Table \ref{tab:band_params}, namely the linear-$k_\pm$ spin--orbit mixing proportional to $R$ and the leading $k_\pm k_z$ mixing proportional to $v$. This yields
\begin{equation}
    \mathcal{T}(\mathbf k)\approx
    \begin{pmatrix}
        -iR\,k_- & \eta(\mathbf k)\\[2pt]
        \eta^*(\mathbf k) & iR\,k_+
    \end{pmatrix},
    \quad
    \eta(\mathbf k)=-v\,k_-k_z,
    \label{eq:T_simplified}
\end{equation}
which is consistent with the full expressions in Eqs.~\eqref{Eq :Tau} and \eqref{Eq :eta_gamma_k} in the limit $w_1,w_2,b,s,t\to 0$. Substituting Eq.~\eqref{eq:T_simplified} into the Löwdin expansion and discarding subleading terms in the $\sigma_x$, $\sigma_z$ channels that are suppressed by $\Delta_1$, we obtain
\begin{align}
    \mathcal{T}\mathcal{T}^\dagger
    =&
    \Bigl(R^2k_\perp^2+|v|^2k_\perp^2k_z^2\Bigr)\,\mathbb{I}_{2\times2} \nonumber
    \\
    &+
    \begin{pmatrix}
    0 & \;2\,iRv\,k_-^2k_z\\[2pt]
    -2\,iRv^*\,k_+^2k_z & 0
    \end{pmatrix}.
    \label{eq:TTdagger}
\end{align}
The first term, being proportional to the identity, merely renormalizes $d_0(\mathbf k)$ and thus leaves the quantum geometry unaffected. In contrast, the second (traceless) term generates a new pseudospin component into the effective Hamiltonian $\mathcal{H}_{45}^{\mathrm{eff}}$. This off-diagonal contribution is essential, as it represents the ``geometric dressing'' of the low-energy manifold, providing the necessary $k$-space non-coplanarity to generate non-zero Christoffel symbols.

Writing the Löwdin downfolded effective Hamiltonian as $\mathcal{H}_{45}^{\mathrm{eff}}=d_0^{\mathrm{eff}}\,\mathbb{I}_{2\times2}+\mathbf d^{\mathrm{eff}}\cdot\boldsymbol\sigma$, and restricting ourselves to the $k_x-k_z$ plane (we shall set henceforth $k_y=0$) we obtain in the simplified $H_{4}-H_5$ block (setting $t=w_1=0$ for the bare block as in Table~\ref{tab:band_params})
\begin{align}
    d_x^{\mathrm{eff}}(\mathbf k)&\simeq -2\Delta_1+2u\,k_z^2,
    \\
    d_y^{\mathrm{eff}}(\mathbf k)&\simeq -\frac{2Rv}{\Delta_2}\,k_x^2 k_z
    \quad\Bigl(\text{for real }v\Bigr),
    \label{eq:dy_eff}
    \\
    d_z^{\mathrm{eff}}(\mathbf k)&\simeq -S_1\,k_z, \label{Eq : H45eff_Pauli_vector}
\end{align}
up to an overall scalar shift in $d_0^{\mathrm{eff}}$. The crucial result is Eq.~\eqref{eq:dy_eff}: downfolding the $H_6,H_{6'}$ sector generates a term
$\propto k_x^2 k_z$, which is {\it odd} in $k_z$ and {\it even} in $k_x$, that was absent in the isolated $\mathcal{H}_{45}$ block. 
This term creates $k_x$-dependent quantum geometry and activates finite Christoffel symbols.

To evaluate the Christoffel symbols analytically, we consider the effective Pauli vector of Eq.~\eqref{Eq : H45eff_Pauli_vector} in the limit $u=0$. Substituting these components into the two-level quantum metric formula in Eq.~\eqref{eq:metric_two_level}, we obtain the metric tensor in the $(k_x,k_z)$ plane as
\begin{align}
    g_{xx}(k_x,k_z)
    &=
    \frac{\alpha^2 k_x^2 k_z^2\bigl(S_1^2k_z^2+m^2\bigr)}{\bigl(m^2+k_z^2(S_1^2+\alpha^2 k_x^4)\bigr)^2},
    \\[4pt]
    g_{xz}(k_x,k_z)
    &=
    \frac{\alpha^2 k_x^3 k_z\,m^2}{2\,\bigl(m^2+k_z^2(S_1^2+\alpha^2 k_x^4)\bigr)^2},
    \\[4pt]
    g_{zz}(k_x,k_z)
    &=
    \frac{m^2\bigl(S_1^2+\alpha^2 k_x^4\bigr)}{4\,\bigl(m^2+k_z^2(S_1^2+\alpha^2 k_x^4)\bigr)^2},
    \label{eq: gzz_plot}
\end{align}
where we have defined the effective mass $m=2\Delta_1$ and $\alpha=2Rv/\Delta_2$. As anticipated from symmetry considerations, the metric components $g_{ii}$ remain even under $k_z\to -k_z$, whereas their derivatives (and hence Christoffel symbols) are generically odd functions of $k_z$.
\begin{figure*}[t]
    \includegraphics[width=1\linewidth]{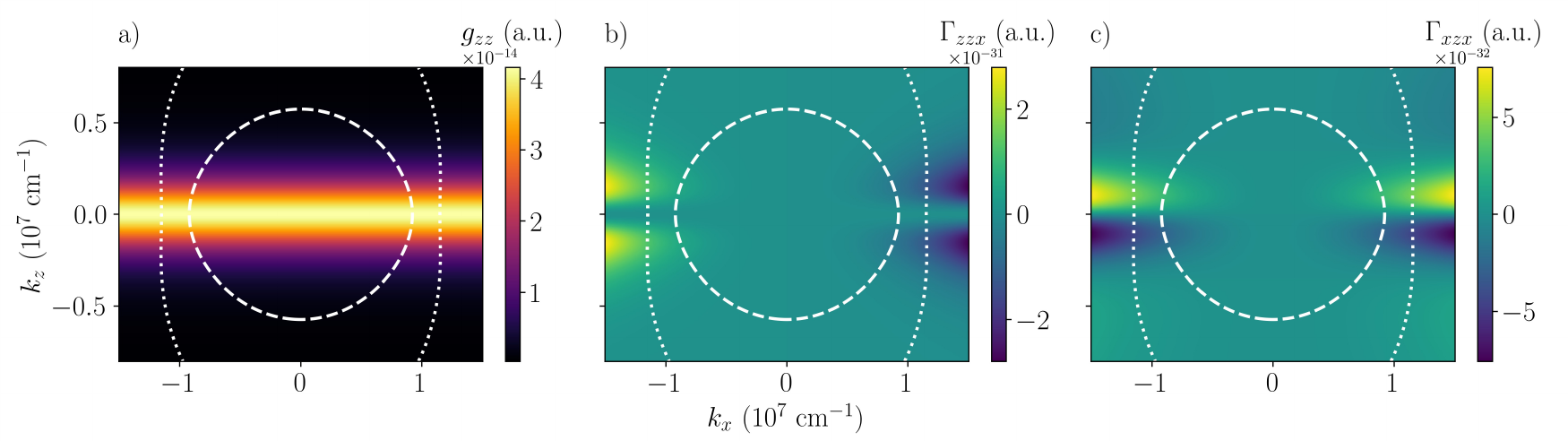}
    \caption{{\bf Geometric origin of the nonreciprocal response.}
    {(a)} Fermi surfaces of the two valence bands in the $(k_x,k_z)$ plane overlaid on a heatmap of the
    energy--normalized quantum metric $\tilde g(\mathbf k)$, which is even under $k_z\rightarrow -k_z$, calculated using Eq. \ref{eq: gzz_plot}.
    {(b)-(c)} The same Fermi surfaces overlaid on a heatmap of the $\Gamma_{zzx}$ and $\Gamma_{xzx}$ Christoffel-symbols, respectively,
    which is odd in $k_z$ and encodes the momentum gradient of the metric.
    While the metric-induced Stark energy shift is symmetric in momentum space, its derivative produces
    asymmetric velocity corrections; in the presence of a magnetic field, this asymmetry leads to
    nonreciprocal transport.
    }
    \label{fig:QM_and_Christoffel_VB}
\end{figure*}
The Christoffel symbols are computed following Eq.~\eqref{eq:Christoffel}. For the eMChA geometry of interest, the relevant components reduce to
\begin{equation}
    \Gamma_{xzx}=\frac{1}{2}\,\partial_{k_z}\tilde g_{xx},
    \qquad
    \Gamma_{zzx}=\partial_{k_z}\tilde g_{xz}-\frac{1}{2}\,\partial_{k_x}\tilde g_{zz}.
    \label{eq:Gamma_reductions}
\end{equation}
By evaluating these derivatives, we obtain the closed-form analytic expressions
\begin{widetext}
\begin{align}
    \Gamma_{xzx}(k_x,k_z)
    &=
    \frac{\alpha^2 k_x^2 k_z\,
    \Bigl(
    2m^4+m^2S_1^2k_z^2-3m^2\alpha^2 k_x^4 k_z^2
    -S_1^2(S_1^2+\alpha^2 k_x^4)k_z^4
    \Bigr)}
    {4\Bigl(m^2+k_z^2(S_1^2+\alpha^2 k_x^4)\Bigr)^{7/2}}
    \label{eq:Gamma_xzx_analytic}
    \\[8pt]
    \Gamma_{zzx}(k_x,k_z)
    &=
    -\frac{5\,\alpha^2\,k_x^3\,k_z^2\,m^2\bigl(S_1^2+\alpha^2 k_x^4\bigr)}
    {8\Bigl(m^2+k_z^2(S_1^2+\alpha^2 k_x^4)\Bigr)^{7/2}}.
    \label{eq:Gamma_zzx_analytic}
\end{align}
\end{widetext}

Both symbols are non-vanishing for generic $(k_x, k_z)$ coordinates within the Brillouin zone. Fig. \ref{fig:QM_and_Christoffel_VB} shows a heatmap of the metric in Eq.~\eqref{eq: gzz_plot} together with the symbols in Eqs.~\eqref{eq:Gamma_xzx_analytic}\eqref{eq:Gamma_zzx_analytic}. Crucially, the latter are odd under $k_z \to -k_z$, as evidenced by the explicit factor of $k_z$ in $\Gamma_{xzx}$ and the resulting parity of the derivatives in $\Gamma_{zzx}$. This odd parity is the defining feature of the QMD. This result contrasts sharply with the isolated $H_4-H_5$ manifold, where the metric $\tilde{g}_{ab}$ lacks the $k_x$ dependence, leading to $\Gamma_{xzx} = \Gamma_{zzx} = 0$.

In summary, Eqs.~\eqref{eq:gzz_45} and \eqref{eq:Gamma_vanish_45} demonstrate that a finite quantum metric is not, by itself, sufficient to generate eMChA. Although the $H_{4}$--$H_{5}$ valence-band block exhibits a non-zero, even-in-$\mathbf k$ metric, its momentum-space dipoles vanish identically, thereby precluding any rectifying contribution to the
current. The emergence of non-reciprocal transport requires the inclusion of the full valence-band structure, in particular the Weyl sector formed by the $H_{6}$ and $H_{6'}$ bands and their spin--orbit-induced hybridization, which generates finite Christoffel symbols and enables the geodesic mechanism detailed in Section \ref{subsec:Boltzmann}.

\subsection{Chemical-potential scaling of the nonlinear transport tensor in the valence sector}
\label{subsec:analytic-G-tensor}
Following the conduction-band analysis of Ref.~\cite{FontanaPRL2025}, we evaluate the nonlinear magnetochiral transport coefficient, $G_{ijkl}$, within the constant-relaxation-time approximation. In this framework, $G_{ijkl}$ is expressed as a Fermi-surface integral involving the band velocity and the geometric dipole, the latter being encoded in the Christoffel symbols. For the valence sector, we restrict the band index in Eq.~\eqref{Eq:G-tensor} to the two uppermost bands ($H_4$ and $H_5$) and replace the thermal broadening term $-\partial f^n_0/\partial\varepsilon^n$ with $\delta(\varepsilon_n(\mathbf{k})-\mu)$. Here, $\mu$ represents the chemical potential measured from the valence-band maximum.

Near the top of the valence sector, the $H_4$ and $H_5$ bands disperse downward. We approximate these as two parabolic branches, neglecting the ``camelback'' shape, which only contributes to subleading orders to the scaling of $G_{ijkl}$. 
The dispersions are 
\begin{align}
    \varepsilon_{4}(\mathbf{k})
    &\simeq
    -\frac{\hbar^{2}k_{\perp}^{2}}{2m_{\perp,4}}
    -\frac{\hbar^{2}k_{z}^{2}}{2m_{z,4}},
    \\
    \varepsilon_{5}(\mathbf{k})
    &\simeq
    -4\Delta_{1}
    -\frac{\hbar^{2}k_{\perp}^{2}}{2m_{\perp,5}}
    -\frac{\hbar^{2}k_{z}^{2}}{2m_{z,5}}.
    \label{eq:valence_parabolic_two_bands}
\end{align}
We set the energy zero at the maximum of the upper valence band ($\varepsilon_{4}(0)=0$), so that $\mu<0$, with $|\mu|$ increasing as the system is doped deeper into the valence sector. The offset $4\Delta_{1}$ reflects the $k=0$ splitting of the $H_4-H_5$ block. Within this approximation, the band velocity scales as $v_i^{n}(\mathbf{k}) \sim \hbar k_i/m_i$, and is therefore {\it odd} in $\mathbf{k}$, whereas its derivatives reduce to constants $\partial_iv^{n}_j=\delta_{ij}\hbar/m_j$. 

As shown in Sec.~\ref{Lowdin-downfolding}, the Löwdin downfolding of the higher-energy $(H_6,H_6')$ block onto the $(H_4,H_5)$ sector introduces an effective pseudospin component to leading order, as in Eq. \eqref
{eq:dy_eff}. This component generates non-zero Christoffel symbols, given in Eqs.~\eqref{eq:Gamma_xzx_analytic} and \eqref{eq:Gamma_zzx_analytic}. Neglecting higher-order terms in momentum, the Christoffel symbols scale as
\begin{equation}
    \Gamma_{jkl}(\mathbf{k})
    \;\sim\;
    \frac{k_{\perp}^{2}k_{z}}
    {\Bigl(\Delta_{1}^{2}+S_{1}^{2}k_{z}^{2}+\alpha^{2}k_{\perp}^{4}k_{z}^{2}\Bigr)^{7/2}},
    \label{eq:Gamma_scaling_k}
\end{equation}
and are therefore \emph{odd} in $k_z$. Consequently, the integrand defining $G_{ijkl}$ in Eq.~\eqref{Eq:G-tensor} is even in $\mathbf{k}$, yielding a finite contribution to the Fermi-surface average.

Given that the energy zero is set at the valence-band maximum, the chemical potential in the valence regime satisfies $\mu<0$. It is therefore convenient introduce the (positive) energy depth
\begin{equation}
    \varepsilon \equiv |\mu|=-\mu>0 .
    \label{eq:eps_depth_def}
\end{equation}
Within the parabolic approximation, the Fermi momentum scales as $k_F^2\propto \varepsilon$, so that $k_F \sim \sqrt{\varepsilon}$. 

Evaluating the Christoffel symbol in Eq. (\ref{eq:Gamma_scaling_k}) at the Fermi momentum and neglecting the $\alpha^2$ terms in the denominator yields
\begin{equation}
    \Gamma(k_F)
    \;\propto\;
    \frac{k_F^3}{\bigl(\Delta_1^2 + S_1^2\,k_F^2\bigr)^{7/2}}
    =
    \frac{\varepsilon^{3/2}}{\bigl(\Delta_1^2 + s_1\,\varepsilon\bigr)^{7/2}},
    \label{eq:Gamma_scaling_valence}
\end{equation}
where we have defined $s_1\equiv S_1^2$. Since the velocity scales as $v(k_F)\sim k_F\sim \sqrt{\varepsilon}$, while its momentum derivatives are constant, the nonlinear conductivity tensor acquires the crossover form
\begin{equation}
    G(\varepsilon)
    \;\propto\;
    \frac{\varepsilon^{3/2}}{\bigl(\varepsilon+\varepsilon_\Delta\bigr)^2}.
    \label{eq:G_scaling_valence_eps}
\end{equation}
Here, the positive crossover energy scale is set by the band-gap term,
\begin{equation}
    \varepsilon_\Delta \equiv \frac{\Delta_1^2}{s_1},
    \label{eq:epsDelta_def}
\end{equation}
up to numerical factors of order unity.

The Eq.~\eqref{eq:G_scaling_valence_eps} captures two distinct limiting transport regimes,
\begin{align}
    G(\varepsilon\gg \varepsilon_\Delta)
    &\sim
    \varepsilon^{-1/2},
    \label{eq:G_large_eps}
    \\[3pt]
    G(\varepsilon\ll \varepsilon_\Delta)
    &\sim
    \varepsilon^{3/2}.
    \label{eq:G_small_eps}
\end{align}
These limits demonstrate that the gap $\Delta_1$ regularizes the nonlinear response close to the band edge $(\varepsilon\ll \varepsilon_\Delta)$, suppressing the would-be infrared divergence and enforcing a vanishing response as $\varepsilon\rightarrow0$. On the other hand, as the system is doped deeper into the valence sector ($\varepsilon \gg \varepsilon_\Delta$), the influence of the gap becomes progressively irrelevant, and the response recovers the gapless power-law asymptote $G\sim \varepsilon^{-1/2}$.

%
\begin{figure}
    \includegraphics[width=1\linewidth]{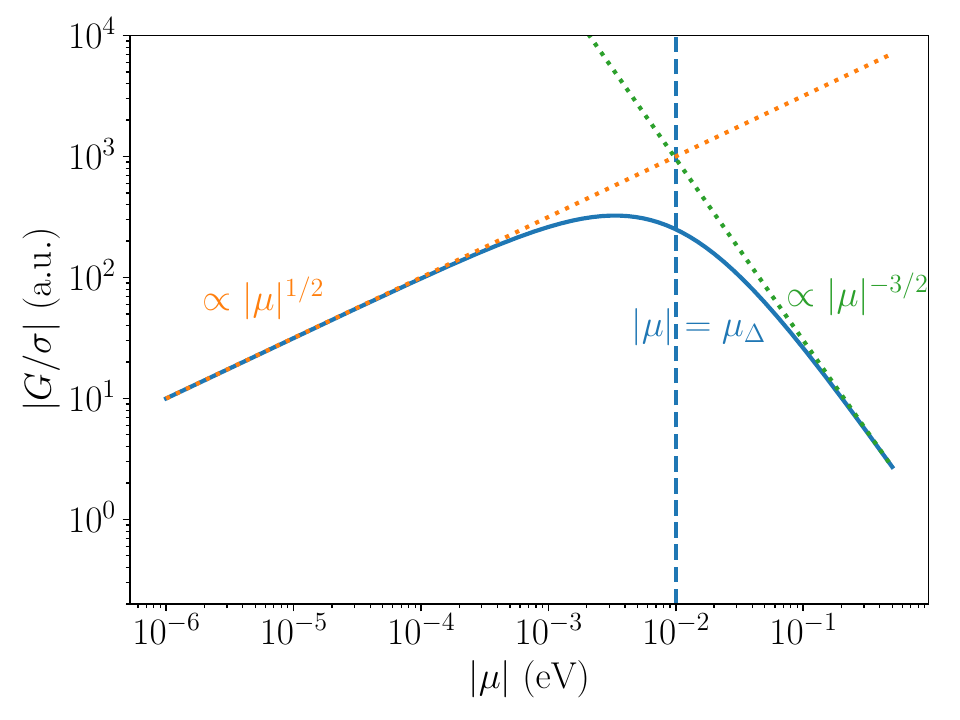}
    \caption{{\bf Scaling behavior of the nonlinear magnetochiral response in the valence band of tellurene.} Log--log plot of the theoretical scaling of the nonlinear magnetochiral conductivity $G$ normalized by $\sigma$ as a function of the energy depth $\varepsilon=|\mu|$ measured from the top of the valence band. Three distinct regimes are identified: (i) a gap-regularized regime near the band edge ($\varepsilon\ll\varepsilon_\Delta$) where $G(\varepsilon)\sim\varepsilon^{3/2}$, (ii) a crossover  regime ($\varepsilon\simeq\varepsilon_\Delta$) where $G(\varepsilon)\sim G_0^{max}$, and (iii) a gapless high-doping regime ($\varepsilon\gg\varepsilon_\Delta$) where $G(\varepsilon)\sim\varepsilon^{-1/2}$.
    }
\label{fig:CrossOver_G_Tensor_VB}
\end{figure}
%

For a 2D parabolic band, the density of states is constant, hence the carrier density scales as $n\propto \varepsilon$ and the Drude conductivity $\sigma(\varepsilon)\propto \varepsilon$. Therefore, the ratio of the nonlinear coefficient to the linear conductivity scales as 
\begin{equation}
    \frac{G(\varepsilon)}{\sigma(\varepsilon)}
    \ \propto\
    \varepsilon^{-3/2}\left(\frac{\varepsilon}{\varepsilon+\varepsilon_\Delta}\right)^2.
    \label{eq:G_over_sigma_valence_eps}
\end{equation}
In the experimentally relevant regime ($\varepsilon\gg\varepsilon_\Delta$),
Eq.~\eqref{eq:G_over_sigma_valence_eps} reduces to the power law
$G/\sigma \sim \varepsilon^{-3/2}$. Conversely, near the valence-band edge ($\varepsilon \ll \varepsilon_\Delta$), the ratio crosses over to the regularized behavior $G/\sigma \sim \varepsilon^{1/2}$. This scaling behaviour of Eq.\eqref{eq:G_over_sigma_valence_eps} is plotted in Fig. \ref{fig:CrossOver_G_Tensor_VB} as a function of the chemical potential for a representative value of the cross-over energy $\varepsilon_\Delta=10^{-2}$.

This analysis establishes that the nonlinear magnetochiral response of the valence sector is governed by the same quantum-geometric dipole mechanism as in the conduction band. However, the response here is enabled by the interband mixing between the low-energy (trivial) and the high-energy (Weyl-node containing) blocks and is characterized by a crossover scale dictated by the intrinsic valence-band splitting $\Delta_1$.

\subsection{Order-of-magnitude estimate of the valence-band eMChA coefficient}
\label{subsec:estimate_valence_gamma}
We provide a back-of-the-envelope estimate for the magnitude of the valence-band eMChA coefficient, following the approach used for the conduction-band estimate in the Supplementary Information of Ref.~\cite{FontanaPRL2025}.
The key experimental observable is the dimensionless ratio between the nonlinear and linear current densities,
\begin{equation}
    \frac{j_{\rm NL}}{j_{\rm L}}=\frac{G\,E\,B}{\sigma}.
    \label{eq:ratio_jNL_jL}
\end{equation}
For the valence-band effective Hamiltonian, the Christoffel symbols scale as $\Gamma\sim \partial \tilde g\propto \alpha^2$, because $g$ is quadratic in interband mixing, and the relevant mixing satisfies $d_y^{\rm eff}\propto \alpha$. Consequently,
\begin{equation}
    G_0^{(v)}\ \propto\ \alpha^2\,\tau^2\,{\cal E}_0\times(\text{band-structure factors}),
    \label{eq:G0val_alpha_scaling}
\end{equation}
which needs to be compared with the conduction-band prefactor controlled by the SOI parameters that enter the Weyl-sector Hamiltonian. This makes transparent why the valence-band signal is naturally suppressed, since the downfolded coupling is \emph{perturbative} in the large scale $\Delta_2$ through $\alpha=2Rv/\Delta_2$.

To obtain a simple numerical estimate for the dimensionless nonlinear ratio, we combine Eqs.~\eqref{eq:ratio_jNL_jL} and \eqref{eq:G_over_sigma_valence_eps} in the experimentally relevant regime (deep enough in the valence band such that $\varepsilon\gg\varepsilon_\Delta$), and parametrize the nonuniversal prefactor by a single amplitude ${\cal A}_v$
\begin{equation}
    \frac{j_{\rm NL}}{j_{\rm L}}
    \;\simeq\;
    {\cal A}_v\,
    \alpha^2\,
    \frac{{\cal E}_0\,E\,B}{\varepsilon^{3/2}},
    \qquad (\varepsilon\gg \varepsilon_\Delta).
    \label{eq:estimate_master}
\end{equation}
with $\tau \simeq 10^{-12}$ s, ${\cal E}_0\simeq 10^9$ V/m, $E\simeq 5\times 10^5$ V/m, $B\simeq 5$ T, $\sigma_{\rm exp}\sim 10^{-6}$ S, we obtain the compact estimate
\begin{widetext}
\begin{equation}
    \frac{j_{\rm NL}}{j_{\rm L}}
    \;\sim\;
    10^{-6}\;
    \left(\frac{{\cal A}_v}{{\cal A}_v^{\rm ref}}\right)\,
    \left(\frac{\alpha}{\alpha_{\rm ref}}\right)^2
    \left(\frac{10~{\rm meV}}{\varepsilon}\right)^{3/2}
    \left(\frac{E}{5\times 10^5~{\rm V/m}}\right)
    \left(\frac{B}{5~{\rm T}}\right),
    \label{eq:valence_numeric_estimate}
\end{equation}
\end{widetext}
where $\alpha_{\rm ref}=2Rv/\Delta_2$ is evaluated from the valence-band parameters (Table~\ref{tab:band_params}), and where ${\cal A}_v^{\rm ref}$ denotes an ${\cal O}(1)$ reference value encoding the remaining band-structure angular factors and effective-mass combinations. Eq.~\eqref{eq:valence_numeric_estimate} shows that a valence-band nonlinear ratio at the level of $10^{-6}$--$10^{-5}$ is fully natural for realistic parameters, i.e., about one order of magnitude (or more) below the conduction-band estimate of $10^{-4}$ \cite{FontanaPRL2025}. This result is consistent with the perturbative origin of $d_y^{\rm eff}\propto 1/\Delta_2$ and the absence of a low-energy Weyl enhancement in the truncated $(H_4,H_5)$ sector.

\section{ Numerical analysis}
\label{sec:numerics}
\begin{figure}
    \includegraphics[scale=0.8]{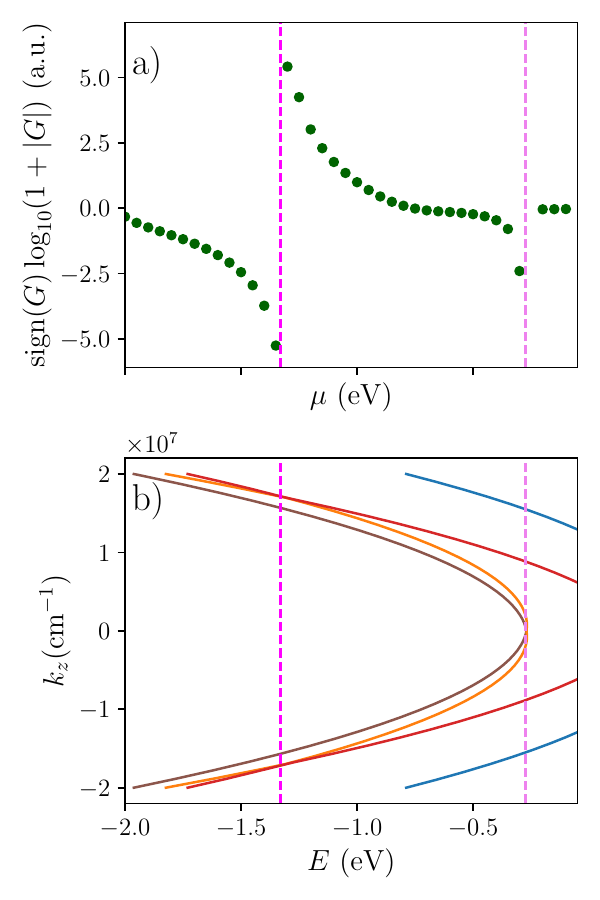}
    \caption{\textbf{Numerical analysis of the $G$-tensor.}
(a) Rescaled value of the $G$-tensor component $G_{zzzy}$ (in arbitrary units)  as a function of the chemical potential $\mu$. (b) Valence bands structure along the $k_z$ direction at $k_x=0$. The dashed lines indicate the energy values of the Weyl nodes in the lower $H_6, H_{6'}$.} 
    \label{fig:Gtensor_Weyls}
\end{figure}
In this Section, we numerically investigate the properties of the valence bands and show how their topology affects the transport properties through non-analytical behaviors of both the density of states and the eMChA tensor. In contrast to the analytical results of Sec.~\ref{sec:Geo-Rect}, which rely on the Löwdin downfolding and further approximations, we exactly account for all bands of the ${\bf k}\cdot {\bf p}$ model presented in Sec.~\ref{sec:model}. This approach allows us to move beyond simple scaling behaviors toward a comprehensive quantitative theory. 

Specifically, we show that the eMChA tensor diverges as the chemical potential approaches the energy of the Weyl nodes. While these values of the chemical potential are not currently accessible experimentally for the valence bands \cite{Chang_2023}, this numerical evidence confirms that the scaling behavior observed of the eMChA tensor for the conduction bands in Ref.~\cite{FontanaPRL2025} is indeed a formal divergence. Furthermore, we establish a connection between the non-trivial topology of the Fermi surfaces at the top of the valence band and the presence of discontinuities in both the density of states and the eMChA tensor derivatives with respect to the chemical potential. 

\begin{figure*}[t]
    \includegraphics[scale=0.6]{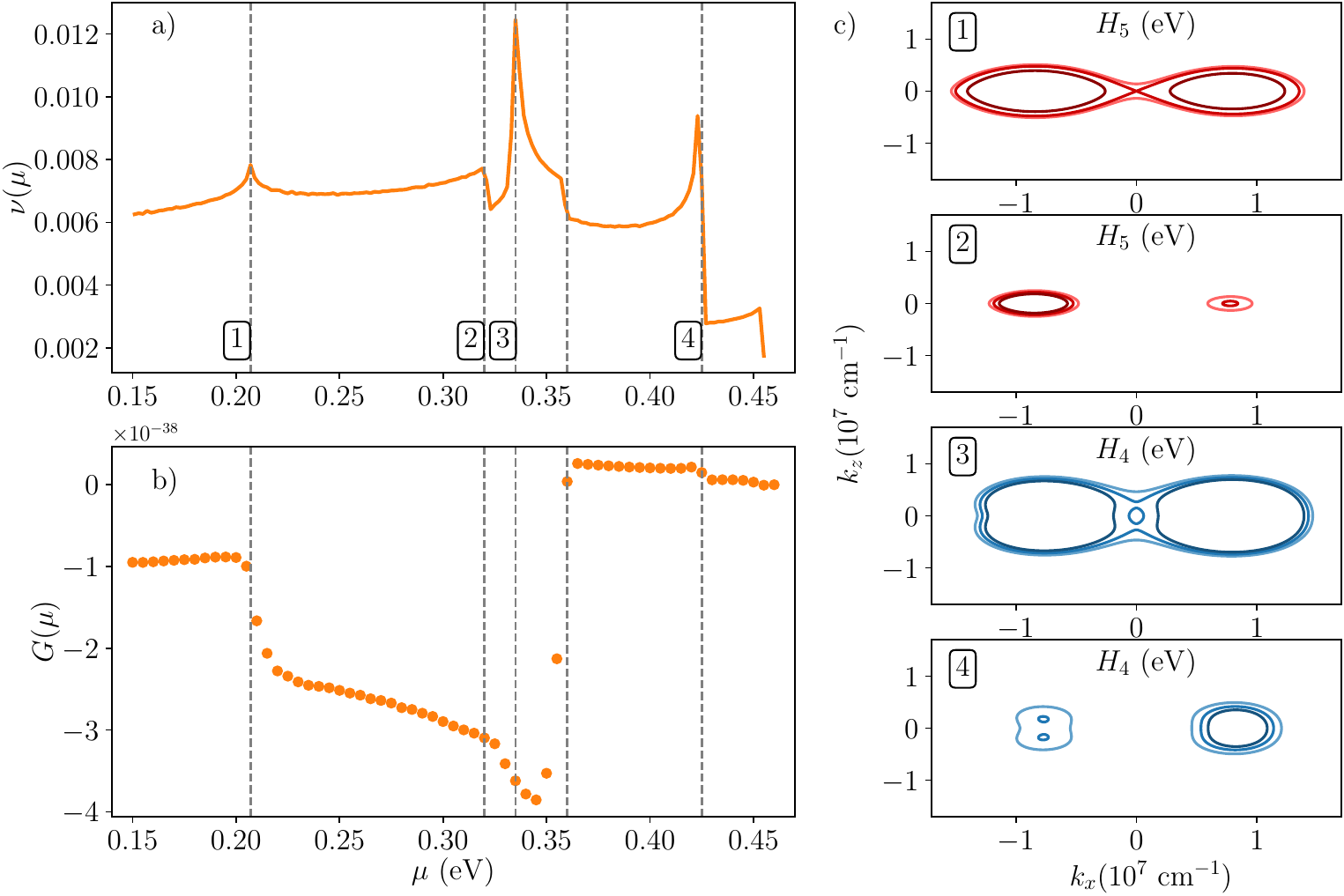}
    \caption{ {\bf Change in Fermi surface topology and associated discontinuities in the G-tensor and density of states.}$\;$(a) The scaling of the density of states $\nu$ in arbitrary units and (b) the $G$ tensor as a function of the chemical potential. 
    A vertical dashed lines is placed where the discontinuity in the derivatives is found. (c) the change of topology in the Fermi surfaces for an energy value equal to the chemical potential where the vertical dashed lines are plotted in (a)-(b). The different Fermi surfaces per plot for the $H_4$ ($H_5$) bands are shaded from lighter to darker blue (red), which corresponds to an increase of the chemical potential. The kink marked by the vertical line at $\mu = 0.37$~eV, to which no right panel has been drawn, is caused by the complete disappearance of the $H_5$ Fermi surface.}
    \label{fig:GDoS}
\end{figure*}
Given the experimental geometry shown in Fig.~\ref{fig:full_bands_Te}, the contribution to the nonreciprocal resistance arises from the $G_{zzzy}$ component of the eMChA tensor defined in Eq.~\eqref{Eq:G-tensor}. To compute this component, we numerically diagonalize the valence-band Hamiltonian in Eq. \eqref{Eq: Hamiltonian-Te} and use finite-difference methods for derivatives across a $k_x$-$k_z$ plane grid at $k_y=0$. This momentum constraint is chosen to match the system geometry, that is, a film in $x$-$z$ plane with negligible extension along the $y$-axis. The summation in momentum space has been performed by replacing the derivative of the equilibrium distribution, namely, the Dirac delta function, with an acceptance region centered at a fixed chemical potential.

In Fig.~\ref{fig:Gtensor_Weyls}, we plot the rescaled value of the $G$-tensor in arbitrary units as a function of the chemical potential measured in $\mathrm{eV}$. We observe that this quantity naturally diverges in proximity of the Weyl nodes, a behavior driven by the divergence of the normalized quantum metric at these points.

In Fig.~\ref{fig:GDoS}(a-b), we analyze both the $G$ tensor and the density of states near the top of the valence band, where only the two uppermost bands contribute to these quantities. The plot shows that the change in Fermi-surface topology results in a discontinuity in the chemical-potential derivatives of both the density of states and the $G$ tensor. This analysis also reveals that the influence of the Weyl nodes within the $H_6, H_{6'}$ sector extends to these observables even when the chemical potential is far from the Weyl energy. Indeed, the hybridization matrix $\mathcal{T}$ is responsible for the camelback shape of the uppermost bands not only along $k_z$, but also along $k_x$, as can be seen in Fig. \ref{fig:GDoS}(c). This quartic-like behavior is absent in the isolated $\mathcal{H}_{45}$ block and is strictly induced by the coupling to the lower bands, consistent with the downfolding mechanism discussed in Sec. \ref{sec:Geo-Rect}. Consequently, the nontrivial topology of the lower valence bands drives the topological transitions in the upper valence bands.

\section{Experimental benchmarking}
\label{sec:Comparison-Experiments}

We now benchmark our theoretical predictions against previously reported second-harmonic transport measurements performed on 2D $\alpha$-tellurene devices, as presented in Refs. \cite{Chang_2023,suarez2025symmetryorigin}. The experimental setup and measurement protocols closely follow those reported in Ref.~\cite{Chang_2023}, and are summarized here for completeness and clarity.

\subsubsection*{Device fabrication and measurement geometry}
In Ref \cite{Chang_2023}, few-layer tellurene flakes were mechanically exfoliated from bulk crystals and transferred onto a Si/SiO$_2$ substrate, with the doped Si layer serving as a global back gate. Standard electron-beam lithography was used to define multi-terminal Hall-bar–like geometries, followed by metal deposition to form ohmic contacts. The crystallographic orientation of each flake was determined prior to contact
fabrication, allowing the longitudinal transport direction to be precisely aligned with the $\hat z$ axis of the tellurene lattice. To isolate the magnetochiral response and minimize spurious Hall contributions, a transverse magnetic field $B_y$ was applied perpendicular to both the current direction and the polar axis. Special care was taken to ensure that the magnetic field alignment was maintained within $1^\circ$ accuracy during field sweeps, a level of precision essential for the clean detection of the nonlinear transport coefficients. 

\subsubsection*{Second-harmonic transport measurements}
Electric magnetochiral anisotropy was probed in Ref. \cite{Chang_2023} using phase-sensitive lock-in techniques. An AC current $I^\omega=I_0\sin(\omega t)$ was injected along the longitudinal ($z$) direction, and the resulting longitudinal voltage $V_{zz}(t)$ was recorded. 
The lock-in amplifier simultaneously measured the first-harmonic $V_{zz}^\omega$ and the second-harmonic $V_{zz}^{2\omega}$ signals. The unidirectional magnetoresistance (UMR) is described by the relation given in Eq. \eqref{Eq:Rgamma}. For ${\bf B}\perp{\bf P}\perp{\bf I}$, the nonlinear voltage components can be expressed by 
\begin{eqnarray}
    V_{zz}^{2\omega}(t)&=&\gamma^\pm R_0 B P I_0^2 \sin^2(\omega t)
    \nonumber\\
    &=&
    \frac{1}{2}\gamma^\pm R_0 B P I_0^2\left[1+\sin\left(2\omega t-\frac{\pi}{2}\right)\right],
    \label{eq: non_linear-V}
\end{eqnarray}
where the lock-in detects the component $V_{zz}^{2\omega}=(1/2)\gamma^\pm R_0 B P I_0^2$ due to its cosine dependence \cite{Chang_2023}. The ordinary magnetoresistance follows the relation $V_{zz}^{\omega}(t)=R_0(1+\beta B^2)I_0\sin(\omega t)$, and the component of the first-harmonic voltage drop detected by the lock-in amplifier is expressed as $V_{zz}^{\omega}=R_0(1+\beta B^2)I_0$ \cite{Chang_2023}. In these experiments, the symmetric part of the longitudinal magnetoresistance, $\overline{R}(B,I,P)\equiv (R(B,I,P)+R(B,-I,P))/2$, exhibited only a minor $\beta{\bf B}^2$ dependence, even at high magnetic fields up to 10 T. Consequently, deviations from linearity in $\overline{R}(B,I,P)$ are negligible within our measurement range and we can safely use $\overline{R}(B,I,P)\approx R_0$. 

The resistance difference between positive and negative current, $\Delta R\equiv R(B,I,P)-R(B,-I,P)$, that encodes the anti-symmetric contribution to the UMR, can be written as
\begin{equation}
    \frac{4 V_{zz}^{2\omega}}{V_{zz}^\omega}=\frac{\Delta R}{R_0}=
    2\gamma^{\pm}{\bf I}\cdot({\bf P}\times{\bf B}).
\end{equation}
In Ref. \cite{Chang_2023}, phase-sensitive measurements were conducted at a temperature of $350$ mK, with magnetic field sweeps up to $\pm10$ T. An AC current of $10\;\mu$A was applied at a frequency of $13.3333$~Hz. Phase-sensitive detection using a lock-in amplifier allowed extraction of the UMR signal. The phases of both the input current and the detection signal are fixed at $0^\circ$. The first-harmonic in-phase ($\hat{x}$) component tracked the linear response, while the second-harmonic quadrature ($\hat{y}$) component, $90^\circ$ out of phase with the input, was used to quantify the UMR.

\begin{figure*}[t]
    \includegraphics[width = 1.0\linewidth]{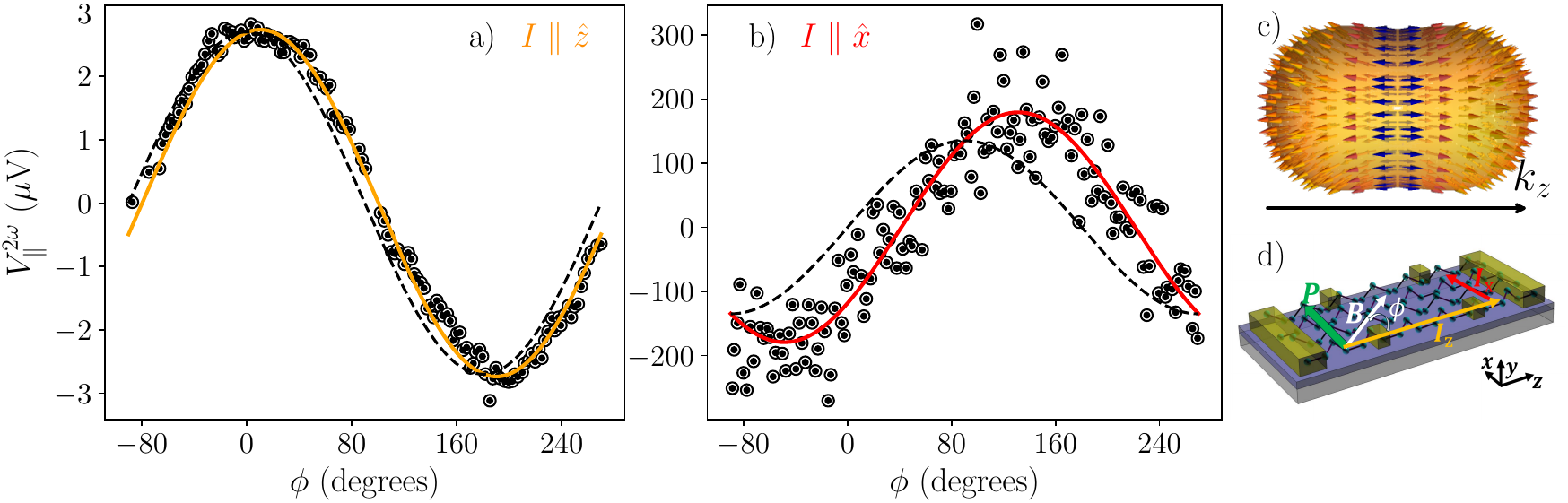}
    \caption{
    {\bf Angular variation of the second harmonic voltage.} 
    (a–b) Longitudinal second-harmonic voltage ${V}^{2\omega}_{\parallel}$ (in $\mu$V) as a function of the in-plane angle $\phi$, measured relative to the $\hat{\mathbf{z}}$ axis [see schematic in (d) for geometry].
    Dotted circles represent experimental data acquired at $T=50$ K and $B=9$ T, 
    extracted from Ref.~\cite{suarez2025symmetryorigin}. 
    The solid orange and red lines are least-squares fits to the form $V^{2\omega}_{zz}(\phi) = A_\chi \cos\phi + A_P \sin\phi$ for $\mathbf{I} \parallel \hat{\mathbf{z}}$, and $V^{2\omega}_{xx}(\phi) = A_\chi \sin\phi + A_P \cos\phi$ for $\mathbf{I} \parallel \hat{\mathbf{x}}$, respectively. The black dashed lines isolate the chiral contribution $\chi$.
    (a) For current $I=1\,\mu$A applied along the $\hat{\mathbf{z}}$ axis, the chiral-type eMChA dominates, resulting in a minimal polar shift.
    (b) For current along the $\hat{\mathbf{x}}$ axis, the chiral and polar contributions are comparable, as evidenced by the significant angular shift.
    (c) Typical Fermi surface of the $H_4$ valence band exhibiting a characteristic radial spin texture.
     The chiral eMChA dominates for ${\bf I}\parallel\hat{\bf z}$ where the magnetic texture is largest due to the current induced spin polarization of holes \cite{shalygin2012current,furukawa2017NComm,FurukawaPhysRevResearch,MurakamiScientificReport,calavalle2022chargetospinconversion}, while the polar eMChA dominates for ${\bf I}\parallel\hat{\bf x}$ due to a lone-pair polarization ${\bf P}\parallel\hat{\bf y}$ \cite{FontanaPRL2025}. 
      (d) Schematic of a two-terminal device and experimental geometry. The current $\mathbf{I}$ is applied either along the $\hat{\mathbf{x}}$ or $\hat{\mathbf{z}}$ directions, while the magnetic field $\mathbf{B}$ rotates in the $\hat{\mathbf{x}}$–$\hat{\mathbf{z}}$ plane at an angle $\phi$ relative to the $\hat{\mathbf{z}}$ axis. This configuration allows for the selective isolation of the chiral ($\mathbf{I} \cdot \mathbf{B}$) and polar ($\mathbf{I} \cdot (\mathbf{P} \times \mathbf{B})$) components of the nonreciprocal response.
    }
    \label{fig:angular_scan_inplane_field}
\end{figure*}

\subsection{Angular dependence $-$ in plane field}
The relative contributions from the two types of mechanisms underlying the UMR are reflected in angular scans with the magnetic field rotated in the $x$-$z$ plane. For a magnetic field $\mathbf{B}=B(\sin\phi\,\hat{\mathbf{x}}+\cos\phi\,\hat{\mathbf{z}})$, one finds for $\mathbf{I}\parallel \hat{\mathbf{z}}$  
\begin{equation}
    \mathbf{I}\cdot\mathbf{B}\ =\ I\, B \cos\phi, \qquad
    \mathbf{I}\cdot(\mathbf{P}\times\mathbf{B})
    = -I\,P_y\,B\,\sin\phi.
    \label{eq:angular_signatures}
\end{equation}
Thus, a dominant $\cos\phi$ harmonic signals the chiral $\mathbf{I}\!\cdot\mathbf{B}$ contribution, whereas a $\sin\phi$ component is the characteristic polar signature $\mathbf{I}\cdot(\mathbf{P}\times\mathbf{B})$. For fields confined to the $x$--$z$ plane, this polar response requires a small but finite $P_y$, corresponding to a weak out-of-plane polar component. For $\mathbf{I}\parallel \hat{\mathbf{x}}$ the roles are reversed and one finds
\begin{equation}
    \mathbf{I}\cdot\mathbf{B}\ =\ I\, B \sin\phi, \qquad
    \mathbf{I}\cdot(\mathbf{P}\times\mathbf{B})
    = I\,P_y\,B\,\cos\phi.
\end{equation}

In Fig.~\ref{fig:angular_scan_inplane_field}(a), we show the angular dependence of the longitudinal second-harmonic voltage $V^{2\omega}_{zz}$ as a function of the in-plane magnetic-field angle $\phi$ measured with respect to the current ${\bf I}\parallel\hat{\bf z}$, with the data extracted from the Supplementary Information of Ref.~\cite{suarez2025symmetryorigin}. 
The dominant cosine component (shown as the black dashed curve), with $A_\chi\approx 2.7\mu$V, reveals the leading chiral contribution proportional to $\mathbf{I}\!\cdot\!\mathbf{B}$ that results from the large spin polarization of holes \cite{Pancharatnam-Berry,Hedgehog} shown in Fig. \ref{fig:angular_scan_inplane_field}(c-d) and produced by the driving current \cite{shalygin2012current,MurakamiScientificReport,furukawa2017NComm,FurukawaPhysRevResearch,calavalle2022chargetospinconversion}, while the statistically significant sine admixture, with $A_P\approx 0.5\mu$V, signals a secondary polar contribution proportional to $\mathbf{I}\!\cdot\!(\mathbf{P}\times\mathbf{B})$ allowed by the nonzero ${\bf P}\parallel\hat{\bf y}$ component \cite{FontanaPRL2025}. The polar contribution accounts for almost $|A_P/A_\chi|\approx 20\%$ of the total signal at $T=50$K, being therefore non-negligible; it is expected to become even more significant at lower temperatures. In Fig.~\ref{fig:angular_scan_inplane_field}(b), we show the same angular dependence of $V^{2\omega}_{xx}$ using data extracted from Ref.~\cite{suarez2025symmetryorigin}, now with the current ${\bf I}\parallel\hat{\bf x}$. The fitted values $A_\chi\approx 135\mu$V and $A_P\approx -120\mu$V (the sign is consistent with $P_y<0$ as reported in \cite{FontanaPRL2025} for $\alpha-$Te) produce a ratio $|A_P/A_\chi|\approx 90\%$, and shows that, in this case, the polar and chiral contributions are comparable in magnitude, which follows naturally from the fact that, for ${\bf I}\parallel\hat{\bf x}$, the spin polarization of holes is much smaller than for ${\bf I}\parallel\hat{\bf z}$ \cite{Hedgehog}.

\subsection{Gate dependence $-$ perpendicular field}
In order to single out the polar contribution to the UMR, the optimal strategy is to set ${\bf B}\perp{\bf I}$, eliminating the chiral contribution, since in this case ${\bf B}\cdot{\bf I}=0$. Accordingly, we fix ${\bf B}\parallel\hat{y}$ and ${\bf I}\parallel\hat{z}$ and analyze the gate voltage dependence of the longitudinal second-harmonic signal measured by the lock-in amplifier. 

A direct comparison between the numerical evaluation of $G/\sigma$ and the analytical scaling requires a consistent mapping between the chemical potential $\mu$ and the experimentally controlled gate voltage $V$. In a two–dimensional system such as tellurene, the density of states, $\nu$, is approximately constant within a parabolic band approximation, so that the carrier density varies linearly with the Fermi energy measured from the band edge, $n  = \nu |\mu|$. On the other hand, electrostatic gating changes the carrier density according to the formula $n = C_g (V-V_0)/e$, where $C_g$ is the gate capacitance per unit area and $V_0$ is the charge–neutrality voltage. 
Combining both relations leads to a linear relation between chemical potential and gate voltage,
\begin{equation}
    |\mu| = \frac{C_g |V-V_0|}{e \,\nu}.
\end{equation}
In the numerical calculations, this correspondence is implemented implicitly by converting the gate–induced carrier density into a Fermi level through the full band dispersion. In contrast, the analytical treatment employs the same proportionality explicitly within the parabolic-band approximation. Although the two procedures differ in detail, both rely on the same physical principle: in 2D, the constant density of states enforces a linear correspondence between $V$ and $\mu$, allowing the numerical and analytical curves of $G(\mu)/\sigma(\mu)$ to be directly compared on the same voltage axis.

\subsection{Comparison between theory and experiments}
\begin{figure}[t]
    \includegraphics[width = 1\linewidth]{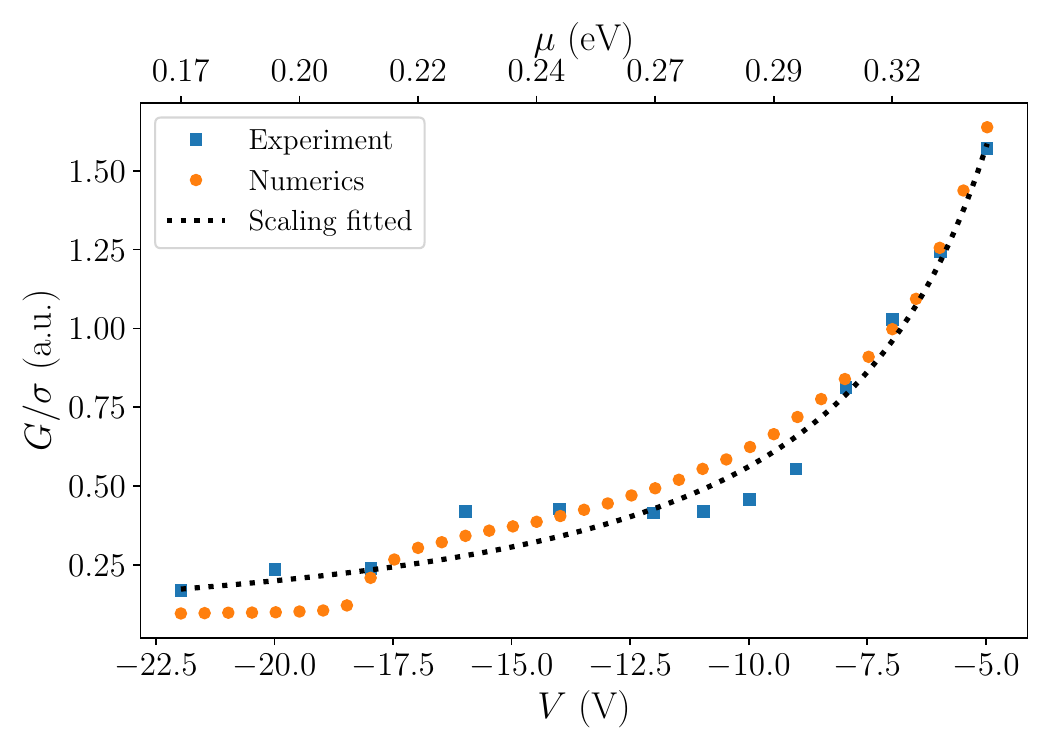}
    \caption{
    {\bf Electric magnetochiral anisotropy in the valence bands of 2D Tellurium.} The blue squares track the backgate voltage dependence of $\Delta R_{zz}/(R_{zz}B_y)$, the orange points show the numerical calculation of $\gamma_{\rm eMChA}$ according to Eq.~\eqref{Eq:Gamma}, for the eMChA tensor component $G_{zzzy}$ and the conductivity $\sigma_{zz}$ given in Eqs.~\eqref{Eq:G-tensor} and \eqref{Eq:sigma_zz}, up to an overall arbitrary rescaling. The dotted line shows the theoretical scaling in Eq.~\eqref{eq:G_over_sigma_valence_eps}, where the proportionality constant and the cross-over energy are determined through fitting the experimental data.}
    \label{fig:rectification.angle}
\end{figure}

Fig.~\ref{fig:rectification.angle} presents the experimentally extracted normalized magnetochiral response, plotted as $\Delta R_{zz}/(R_{zz}B_y)$, as a function of back-gate voltage. These data are recast in terms of the coefficient $\gamma_{\rm eMChA}$ and compared with the theoretical prediction obtained from the QMD formalism developed in Secs.~\ref{sec:Geo-Rect} and \ref{sec:numerics}. We find good agreement between the measured gate dependence of $\gamma_{\rm eMChA}(\mu)$ and the rescaled theoretical expression derived from Eqs.~\eqref{Eq:G-tensor}, \eqref{Eq:Gamma}, and \eqref{Eq:sigma_zz}, up to an overall prefactor that depends on the device geometry and the scattering time, as discussed in Sec.~\ref{subsec:estimate_valence_gamma}.

In particular, the nonmonotonic behavior observed as a function of gate voltage reflects the crossover between distinct valence-band regimes identified theoretically. The suppression of the response near the top of the valence band is consistent with the gap-induced regularization discussed in Sec.~\ref{sec:numerics}, while the enhanced response at larger hole densities arises from the combined influence of QMDs and the concomitant reduction of the longitudinal conductivity. Small deviations from the smooth theoretical trend can be attributed to Landau quantization and Shubnikov--de Haas oscillations at finite magnetic field \cite{shoenberg1984magnetic}, that modulate both the density of states and the scattering time, but do not alter the underlying geometric mechanism.

Overall, these experimental results provide strong evidence that the UMR in the valence bands of tellurene originates from a quantum-geometric mechanism enabled by multiband hybridization. The strong qualitative agreement with our analytical predictions, combined with the quantitative agreement with our numerical calculations, provides compelling support for the theoretical framework developed in this work.

\section{Conclusions}
\label{Conclusions}
In this work, we have investigated the electric magnetochiral anisotropy (eMChA) in the valence-band sector of two-dimensional tellurene, establishing a quantum-geometric mechanism for polar unidirectional magnetotransport in a regime where such an effect would not be expected on the basis of band topology alone. While our previous work demonstrated that eMChA in the conduction band of tellurene is governed by quantum metric dipoles (QMDs) associated with Weyl nodes, the present study reveals a qualitatively different and more subtle origin of nonreciprocal transport for hole carriers.

We have shown that the two uppermost valence bands ($H_4$ and $H_5$), which are topologically trivial and do not host Weyl nodes, possess a finite quantum metric but exhibit identically vanishing QMDs when considered in isolation. Consequently, these bands alone cannot generate eMChA. The emergence of a finite eMChA in the valence band therefore relies on multiband effects. Specifically, Löwdin downfolding of deeper valence bands, separated by a large energy scale $\Delta_2$, generates additional contributions to the effective low-energy Hamiltonian. Crucially, the corresponding hybridization matrix elements are explicitly momentum dependent and scale as $k^2$. These terms enter in the numerator of the Christoffel symbols that define the QMD, while the energy denominator is governed by the large and nearly constant scale $\Delta_2$. 
As a result, the nonlinear response acquires an extra factor of $k^2\sim\mu$ relative to the Weyl-node dominated conduction-band case.

Within a semiclassical Boltzmann framework, we derived analytic expressions for the nonlinear magnetochiral conductivity in the valence regime and established its scaling with chemical potential. It is instructive to contrast this behavior with that of the conduction bands studied in Ref.~\cite{FontanaPRL2025}. In the presence of Weyl nodes near the Fermi level, the QMD is intrinsic to the low-energy bands, resulting in a characteristic nonlinear response scaling of $\gamma(\mu)\sim\mu^{-5/2}$. In contrast, the $H_4-H_5$ valence-band block considered here is topologically trivial and free of Weyl nodes. Accordingly, no intrinsic QMD exists within this truncated description.

This mechanism naturally leads to a modified scaling behavior. Close to the top of the valence band, the response is regularized by the intrinsic valence-band gap, suppressing nonlinear conductivity. Deeper in the valence sector, where $\mu$ is large compared to the gap scale, momentum-dependent hybridization effects dominate, and the normalized response follows a $G/\sigma\sim\mu^{-3/2}$ power law. We provide extensive numerical calculations that quantitatively validate this picture. The transition from the $\mu^{-5/2}$ scaling observed in the conduction band to the $\mu^{-3/2}$ scaling found here is therefore not anomalous, but instead a direct consequence of the momentum structure induced by multiband mixing.

Our results highlight the central role of band hybridization and quantum geometry in nonlinear transport phenomena, demonstrating that unidirectional magnetoresistance is not restricted to topological bands. More broadly, they establish tellurene as a model platform for exploring quantum-geometric rectification across both electron and hole doping regimes.

Finally, we return to the symmetry-based resistivity scaling analysis recently reported in Ref.~\cite{suarez2025symmetryorigin}, which establishes a remarkably clear framework to classify the microscopic origin of eMChA in tellurium through the expansion of the second-harmonic voltage in powers of the longitudinal resistivity
\begin{equation}
    \frac{V^{2\omega}}{(I^\omega)^2}=\cdots+\xi\,\rho+\cdots
    \label{eq:resistivity_expansion}
\end{equation}
In Ref.~\cite{suarez2025symmetryorigin}, the term linear in $\xi\,\rho$ was identified as an intrinsic contribution, distinct from terms governed by scattering processes and extrinsic mechanisms. Our results uncover the microscopic origin of precisely this linear-in-$\rho$ contribution. Within our semiclassical framework, the quantum-geometric eMChA conductivity tensor $G_{ijk\ell}$ scales as $\tau^2$. Since the experimentally accessed second-harmonic voltage is proportional to $\rho^3 G$, and since $\rho \sim 1/\tau$, it follows that $G\sim \rho^{-2}$ and the quantum-geometric contribution to the eMChA enters the resistivity expansion, Eq.~\eqref{eq:resistivity_expansion}, exactly as a term proportional to $\rho$.

This observation establishes a direct link between the phenomenological symmetry analysis of Ref.~\cite{suarez2025symmetryorigin} and the microscopic quantum-geometric mechanism developed in the present work. In particular, it shows that the linear-in-$\rho$ term identified there as intrinsic is not merely symmetry-allowed, but is in fact governed by the momentum-space geometry of the Bloch states through the Christoffel symbols of the energy-weighted quantum metric. The lone-pair polarization of tellurene acts as an internal built-in electric field that activates this geometric channel, converting the even-in-$\mathbf{k}$ Stark renormalization of the bands into an odd-in-current velocity correction under Lorentz deflection. 

Looking ahead, the sensitivity of eMChA to band structure, polarization, and external fields suggests promising opportunities for engineering nonreciprocal responses via electrostatic gating, strain, or heterostructure design. Extending this framework to time-dependent driving, stronger magnetic fields, or other low-dimensional polar materials may uncover even richer connections between multiband quantum geometry and nonlinear electronic transport.

\begin{acknowledgments}
V.V. acknowledges financial support of PNRR MUR project PE0000023-NQSTI and PRIN 2022 (Prot. 20228YCYY7). This work is supported by the Brazilian funding agencies FAPERJ (grant numbers E-26/210.100/2023, E-26/204.003/2024, and E-26/210.781/2025) and CNPq (grant numbers 313059/2020-9 and 442072/2023-6). 
\end{acknowledgments}

\appendix
\widetext 

\section{Tensor Analysis and the eMChA parameter}

\subsection{\label{Ap: GtoGamma}From the nonlinear current response to the SHG tensor {$\gamma$}}
We start from the constitutive expansion of the AC current density 
\begin{equation}
    j_i(t)=\sigma_{ij}E_j(t)+\sigma^{(H)}_{ijk}\,E_j(t)B_k+G_{ijk\ell}\,E_j(t)E_k(t)B_\ell,
    \label{eq:j_expansion}
\end{equation}
where $G_{ijkl}$ is symmetric under $j\leftrightarrow k$, and we retain only terms up to order $E^2B$. Next, we assume a single-frequency drive at $\omega$ and a static magnetic field $B$,
\begin{equation}
    E_i(t)=\Re\!\left[E_i^\omega e^{-i\omega t}+E_i^{2\omega} e^{-i2\omega t}\right],
    \qquad
    B=\text{const}.
\end{equation}
The quadratic term $E_j(t)E_k(t)$ contains a $2\omega$ component that includes
\begin{equation}
    E_j(t)E_k(t)\supset \frac{1}{2}\,E_j^\omega E_k^\omega\,e^{-i2\omega t} + \text{c.c.},
    \label{eq:EE_2w}
\end{equation}
where the factor $1/2$ is the standard consequence of taking the real part.

In a typical second-harmonic transport measurement with an injected current of frequency $\omega$, the source fixes the current at $\omega$ but does not inject current at $2\omega$. One therefore imposes the open-circuit condition at the second harmonic
\begin{equation}
    j_i^{2\omega}=0.
    \label{eq:j2w_zero}
\end{equation}
At order $E^2B$, the Hall term $\sigma^{(H)}_{ijk}E_jB_k$ produces a response only at the same frequency as $E$ (being linear in $E$), and thus does not generate a $2\omega$ component from a purely $\omega$-frequency drive. Consequently, the $2\omega$ component of Eq.~\eqref{eq:j_expansion} is
\begin{equation}
    j_i^{2\omega}
    =
    \sigma_{ij}E_j^{2\omega}
    +
    \frac{1}{2}\,G_{ijk\ell}\,E_j^\omega E_k^\omega\,B_\ell
    \quad
    \stackrel{\eqref{eq:j2w_zero}}{=}\quad 0.
    \label{eq:j2w_balance}
\end{equation}
%

\subsubsection*{First tensor inversion: $\rho=\sigma^{-1}$ and $E^{2\omega}$ in terms of $E^\omega$}

As usual, we define the (linear) resistivity tensor as the inverse of $\sigma$,
\begin{equation}
    \rho_{im}\sigma_{mj}=\delta_{ij}, \qquad \rho\equiv \sigma^{-1}.
    \label{eq:rho_def}
\end{equation}
Left-multiplying Eq.~\eqref{eq:j2w_balance} by $\rho_{ai}$ gives
\begin{equation}
    E_a^{2\omega}
    =
    -\frac{1}{2}\,\rho_{ai}\,G_{ijk\ell}\,E_j^\omega E_k^\omega\,B_\ell.
    \label{eq:E2w_in_terms_of_Ew}
\end{equation}
Eq. \eqref{eq:E2w_in_terms_of_Ew} is the cleanest place where the output-channel inversion ($\sigma \mapsto \rho$) appears explicitly.

\subsubsection*{Second inversion: $E^\omega$ in terms of $j^\omega$ and emergence of $\gamma_{ijk\ell}$}

At the fundamental frequency, keeping only the \emph{linear} relation between
$j^\omega$ and $E^\omega$ is sufficient to obtain $E^{2\omega}$ correctly to order $Bj^2$,
\begin{equation}
    j_i^\omega=\sigma_{ij}E_j^\omega + \mathcal{O}(B E^\omega),
    \qquad\Rightarrow\qquad
    E_j^\omega=\rho_{jm}\,j_m^\omega + \mathcal{O}(B j^\omega).
    \label{eq:Ew_from_jw}
\end{equation}
Any $B$-correction to $E^\omega$ would generate only $\mathcal{O}(B^2j^2)$ terms, beyond the order of interest. Substituting Eq. \eqref{eq:Ew_from_jw} into Eq. \eqref{eq:E2w_in_terms_of_Ew} yields
\begin{align}
    E_a^{2\omega}
    &=
    -\frac{1}{2}\,\rho_{ai}\,G_{ijk\ell}\,
    (\rho_{jm}j_m^\omega)(\rho_{kn}j_n^\omega)\,B_\ell
    \nonumber\\[4pt]
    &\equiv
    \gamma_{a m n \ell}\,j_m^\omega j_n^\omega\,B_\ell,
    \label{eq:gamma_def}
\end{align}
so that the second-harmonic generation tensor is
\begin{equation}
    \gamma^{(E)}_{a m n \ell}
    =
    -\frac{1}{2}\,
    \rho_{ai}\,G_{ijk\ell}\,\rho_{jm}\rho_{kn}.
    \label{eq:gamma_in_terms_of_G}
\end{equation}
This corresponds to the full tensor inversion: one inversion in the output index ($\rho_{ai}$) and two inversions in the \emph{input indices} ($\rho_{jm}$ and $\rho_{kn}$), converting the $E^\omega E^\omega B$ nonlinearity into a $j^\omega j^\omega B$ response for the generated field. This object also characterizes the electric field $E^{2\omega}$ generated under open-circuit conditions at the second harmonic. In an isotropic system, $\gamma^{(E)}\sim G/\sigma^{3}$.

\subsection{\label{Ap: RandGamma}Experimental eMChA coefficient and normalization by the linear resistance}

The experimentally relevant eMChA coefficient is defined through the nonlinear resistance expansion
\begin{equation}
    R(I,B,P)=R_0\Big[1+\beta B^2+\gamma_{\rm eMChA}\,
    \mathbf{I}\cdot(\mathbf{P}\times\mathbf{B})\Big],
    \label{eq:R_expansion}
\end{equation}
and is extracted from the ratio between the second-harmonic and first-harmonic voltages,
\begin{equation}
    \frac{4V_{zz}^{2\omega}}{V_{zz}^{\omega}}
    =
    \frac{\Delta R}{R_0}
    =
    2\,\gamma_{\rm eMChA}\,
    \mathbf{I}\cdot(\mathbf{P}\times\mathbf{B}).
    \label{eq:ratio_exp}
\end{equation}
Since both voltages are measured along the same path, $V^\omega\propto E^\omega$ and $V^{2\omega}\propto E^{2\omega}$, the geometric factors cancel in the ratio,
\begin{equation}
    \frac{V^{2\omega}}{V^\omega}
    =
    \frac{E^{2\omega}}{E^\omega}.
    \label{eq:V_to_E_ratio}
\end{equation}
From the field-based second-harmonic generation tensor derived above,
\begin{equation}
    E^{2\omega}
    \sim
    \gamma^{(E)}\,j^2 B,
    \qquad
    \gamma^{(E)} \sim \frac{G}{\sigma^3},
    \label{eq:E2w_scaling}
\end{equation}
while the fundamental electric field follows from linear response,
\begin{equation}
    E^\omega = \rho j = \frac{j}{\sigma}.
    \label{eq:Ew_scaling}
\end{equation}
Combining Eqs.~\eqref{eq:V_to_E_ratio}–\eqref{eq:Ew_scaling} yields
\begin{equation}
    \frac{V^{2\omega}}{V^\omega}
    =
    \frac{(G/\sigma^3)\,j^2 B}{j/\sigma}
    =
    \frac{G}{\sigma^2}\,j\,B.
    \label{eq:V_ratio_intermediate}
\end{equation}
Eq.~\eqref{eq:V_ratio_intermediate} is an exact consequence of the open-circuit condition at $2\omega$ and involves the \emph{current density} $j$.
To connect with the experimentally defined coefficient $\gamma_{\rm eMChA}$,
we express the current density and the linear resistance in terms of the total current $I$,
\begin{equation}
    j = \frac{I}{A},
    \qquad
    R_0 = \frac{L}{A\sigma},
    \label{eq:j_and_R0}
\end{equation}
where $A$ is the cross-sectional area and $L$ the voltage probe separation.
Substituting Eq.~\eqref{eq:j_and_R0} into Eq.~\eqref{eq:V_ratio_intermediate} gives
\begin{equation}
    \frac{V^{2\omega}}{V^\omega}
    =
    \frac{G}{\sigma^2}\,
    \frac{I}{A}\,B
    =
    \left(\frac{G}{\sigma}\right)
    \left(\frac{I B}{A\sigma}\right).
    \label{eq:V_ratio_geometry}
\end{equation}
Recognizing that $1/(A\sigma)\propto R_0/L$, and absorbing the purely geometric
factor $L$ into the definition of $\gamma_{\rm eMChA}$, Eq.~\eqref{eq:ratio_exp}
is recovered with the scaling
\begin{equation}
    \gamma_{\rm eMChA} \sim \frac{G}{\sigma}.
    \label{eq:gamma_scalar_scaling}
\end{equation}
The origin of the scaling $\gamma_{\rm eMChA}\sim G/\sigma$ can be understood by tracking how the three powers of the linear conductivity appearing in the field-based SHG tensor, $\gamma^{(E)}\sim G/\sigma^{3}$, are redistributed through experimental normalization. One power of $\sigma$ is removed when forming the ratio between second- and first-harmonic fields, $E^{2\omega}/E^\omega$, since $E^\omega\sim j/\sigma$. A second power is absorbed when converting the current density $j$ into the total current $I$
and expressing $1/(A\sigma)$ in terms of the linear resistance $R_0=L/(A\sigma)$, that appears outside brackets in Eq.~\eqref{eq:R_expansion}.
After these normalizations, only a single inverse power of the conductivity remains, yielding the experimentally relevant scaling $\gamma_{\rm eMChA}\sim G/\sigma$.

\section{\label{Ap: Lowdin_Appendix}Effective L\"owdin Hamiltonian}

\begin{figure*}
\includegraphics[scale=0.7]{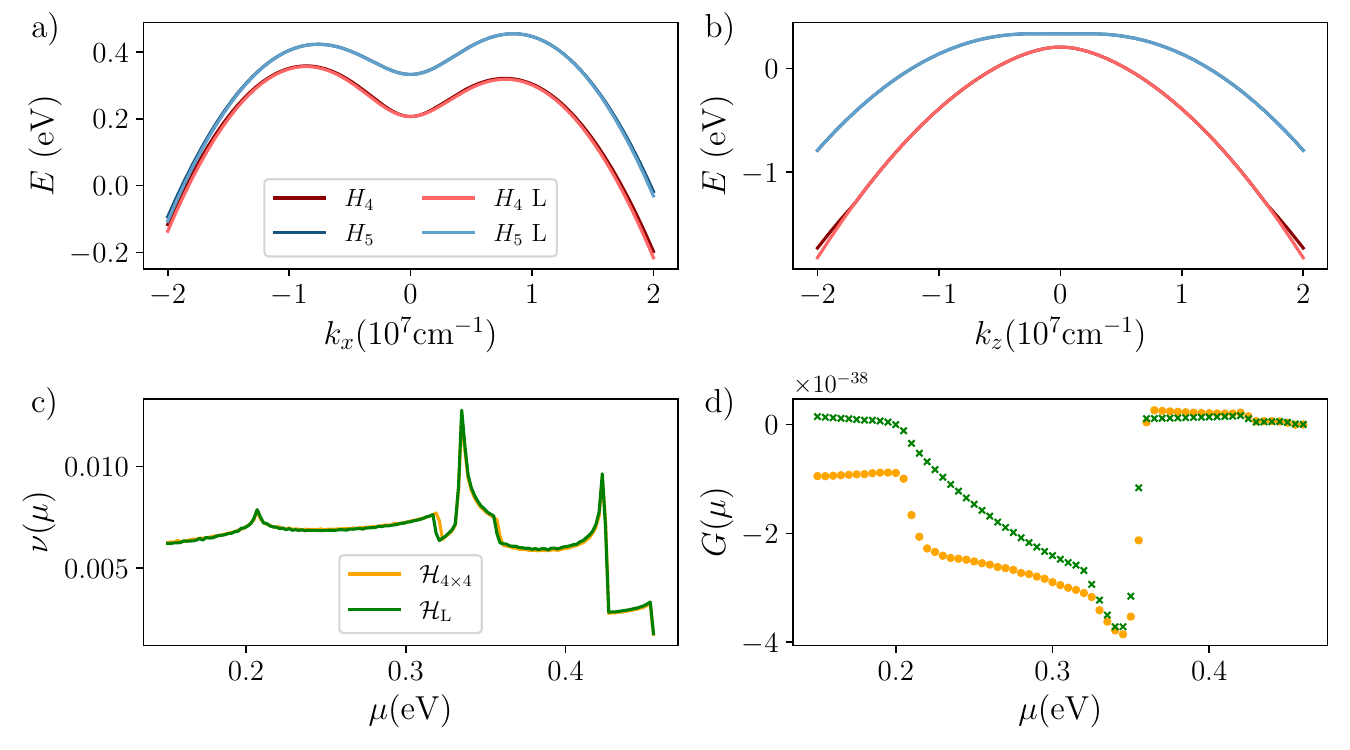}
\caption{Comparison of the dispersion relation along $k_z$ (a) and $k_x$ (b), (c) density of states $\nu$ and (d) $G_{zzzy}$ tensor from  $\mathcal{H}_{4\times 4}$ and $\mathcal{H}_\mathrm{L}$. }
\label{fig:Lowdin}
\end{figure*}

To derive an effective $2\times 2$ Hamiltonian from the full valence band one \eqref{Eq: Hamiltonian-Te}, we start from the corresponding eigenvalue problem, which couples the low- and high-energy subspaces
\begin{equation}
\begin{pmatrix}
    {\cal H}_{45} & {\cal T} \\
    {\cal T}^\dagger & {\cal H}_{66^\prime}
\end{pmatrix}
\begin{pmatrix}
    \Psi \\
    \Phi
\end{pmatrix}
=
{\cal E}
\begin{pmatrix}
    \Psi \\
    \Phi
\end{pmatrix}.
\end{equation}
Here the low-energy components are $\Psi=(\psi_4,\psi_5)^T$, corresponding to $H_4$ and $H_5$ representations, while the high-energy components are $\Phi=(\phi_6,\phi_{6^\prime})^T$, associated with $H_6$ and $H_6'$. From the second row of the eigenvalue equation,
\begin{equation}
    {\cal T}^\dagger \Psi + {\cal H}_{66^\prime} \Phi = {\cal E} \Phi,
\end{equation}
the high-energy components can be expressed in terms of the low-energy ones as
\begin{equation}
    \Phi=({\cal E}-{\cal H}_{66^\prime})^{-1}{\cal T}^\dagger \Psi.
\end{equation}
Substituting this expression into the first row yields
\begin{eqnarray}
    {\cal H}_{45}\Psi+{\cal T}\Phi&=&{\cal E}\Psi,\nonumber\\
    \left[{\cal H}_{45}+{\cal T}({\cal E}-{\cal H}_{66^\prime})^{-1}{\cal T}^\dagger\right]\Psi&=&{\cal E}\Psi,\label{Lowdin-full}.
\end{eqnarray}
To retain a quantitatively good description around the Fermi surface  $\epsilon_0\approx 0.455$~eV, we can perform a Taylor expansion of the matrix inverse $({\cal E}-{\cal H}_{66^\prime})^{-1} = (\epsilon_0 \mathbb{I}- {\cal H}_{66'})^{-1} - (\epsilon_0\mathbb{I}-{\cal E})(\epsilon_0 \mathbb{I}- {\cal H}_{66'})^{-2}+\mathcal{O}({\cal E}^2)$ in powers of ${\cal E}$ around this energy. Substituting the expansion and rearranging terms we find 
\begin{equation}
      \left[{\cal H}_{45}+{\cal T}  \mathcal{Q}{\cal T}^\dagger\right]\Psi=\mathcal{S}{\cal E}\Psi,
\end{equation}
 where
\begin{equation}
    \mathcal{Q}=(\epsilon_0 \mathbb{I}- {\cal H}_{66'})^{-1}+\epsilon_0(\epsilon_0 \mathbb{I}- {\cal H}_{66'})^{-2},\quad{\cal S}=1+{\cal T}(\epsilon_0\mathbb{I}-{\cal H}_{66^\prime})^{-2}{\cal T}^\dagger. 
\end{equation}
Within L\"owdin's perturbation theory \cite{Lowdin}, this defines an effective eigenvalue problem with Hamiltonian
\begin{eqnarray}
    \mathcal{H}_{L}=\mathcal{S}^{-1}\left[{\cal H}_{45}-{\cal T}\mathcal{Q}{\cal T}^\dagger\right].
    \label{eq:H_Lowdin}
\end{eqnarray}
In Fig.~\ref{fig:Lowdin}, we compare the band energies, the density of states $\nu$ and the $G_{zzzy}$ tensor near the Fermi surface for both the full \eqref{Eq: Hamiltonian-Te} and the  L\"owdin Hamiltonian \eqref{eq:H_Lowdin}. 
We notice that, although the band dispersion and the density of states are in a strong quantitative agreement, the $G_{zzzy}$, which requires the full eigensystem and its derivatives, differs significantly in the $\mu$ region considered for the comparison to the experiments in Sec.~\ref{sec:Comparison-Experiments}.

\twocolumngrid

\bibliography{biblio}

\end{document}